\documentclass[aps,prl,superscriptaddress]{revtex4-2}
\setcitestyle{super}
\usepackage{amsmath}	
\usepackage{graphicx}	
\usepackage{color,soul}
\usepackage{gensymb}
\usepackage{braket}
\usepackage{hyperref}
\usepackage{upgreek}

\begin{document}
\Large

\title{\Large Resonant band hybridization in alloyed transition metal dichalcogenide heterobilayers}

\author{Alessandro Catanzaro}
\affiliation{Department of Physics and Astronomy, The University of Sheffield, Sheffield S3 7RH, UK}

\author{Armando Genco}
\email{armando.genco@polimi.it}
\affiliation{Department of Physics and Astronomy, The University of Sheffield, Sheffield S3 7RH, UK}
\affiliation{{Dipartimento di Fisica, Politecnico di Milano, Piazza Leonardo da Vinci, 32, Milano, 20133, Italy}}

\author{Charalambos Louca}
\affiliation{Department of Physics and Astronomy, The University of Sheffield, Sheffield S3 7RH, UK}
\affiliation{{Dipartimento di Fisica, Politecnico di Milano, Piazza Leonardo da Vinci, 32, Milano, 20133, Italy}}

\author{David A. Ruiz-Tijerina}
\affiliation{Departamento de F\'isica Qu\'imica, Instituto de F\'isica, Universidad Nacional Aut\'onoma de M\'exico, Ciudad de M\'exico, C.P.\ 04510, M\'exico}

\author{Daniel J. Gillard}
\affiliation{Department of Physics and Astronomy, The University of Sheffield, Sheffield S3 7RH, UK}

\author{Luca Sortino}
\affiliation{Department of Physics and Astronomy, The University of Sheffield, Sheffield S3 7RH, UK}
\affiliation{Chair in Hybrid Nanosystems, Nanoinstitute Munich, Faculty of Physics, Ludwig-Maximilians-Universität München, Munich, Germany}

\author{Aleksey Kozikov}
\affiliation{Department of Physics and Astronomy, University of Manchester, Manchester, M13 9PL, UK}
\affiliation{School of Mathematics, Statistics and Physics, Newcastle University, Newcastle upon Tyne, NE1 7RU, UK}

\author{Evgeny M. Alexeev}
\affiliation{Department of Physics and Astronomy, The University of Sheffield, Sheffield S3 7RH, UK}
 \affiliation{Cambridge Graphene Centre, University of Cambridge, 9 J. J. Thomson Avenue, Cambridge, CB3 0FA, UK}

\author{Riccardo Pisoni}
\affiliation{Solid State Physics Laboratory, ETH Zurich, Zurich, Switzerland}

\author{Lee Hague}
\affiliation{National Graphene Institute, University  of  Manchester,  Manchester  M13  9PL,  UK}

\author{Kenji Watanabe}
\affiliation{Research Center for Electronic and Optical Materials, National Institute for Materials Science, 1-1 Namiki, Tsukuba 305-0044, Japan}

\author{Takashi Taniguchi}
\affiliation{Research Center for Materials Nanoarchitectonics, National Institute for Materials Science,  1-1 Namiki, Tsukuba 305-0044, Japan}

\author{Klauss Ensslin}
\affiliation{Solid State Physics Laboratory, ETH Zurich, Zurich, Switzerland}

\author{Kostya S. Novoselov}
\affiliation{Institute for Functional Intelligent Materials, National University of Singapore, Singapore, 117546, Singapore}

\author{Vladimir Fal'ko}
\affiliation{Department of Physics and Astronomy, University of Manchester, Manchester, M13 9PL, UK}

\author{Alexander I. Tartakovskii}
\email{a.tartakovskii@sheffield.ac.uk}
\affiliation{Department of Physics and Astronomy, The University of Sheffield, Sheffield S3 7RH, UK}

\begin{abstract}

\end{abstract}

\maketitle

\begin{center}
\textbf{\textit{In memory of Alessandro Catanzaro}}
\end{center}
\vspace{10px}

\textbf{Bandstructure engineering using alloying is widely utilised for achieving optimised performance in modern semiconductor devices.  While alloying has been studied in monolayer transition metal dichalcogenides, its application in van der Waals heterostructures built from atomically thin layers is largely unexplored. Here, we fabricate heterobilayers made from monolayers of WSe$_2$ (or MoSe$_2$) and Mo$_x$W$_{1-x}$Se$_2$ alloy and observe nontrivial tuning of the resultant bandstructure as a function of concentration $x$. We monitor this evolution by measuring the energy of photoluminescence (PL) of the interlayer exciton (IX) composed of an electron and hole residing in different monolayers. In Mo$_x$W$_{1-x}$Se$_2$/WSe$_2$, we observe a strong IX energy shift of $\approx$100 meV for $x$ varied from 1 to 0.6. However, for $x<0.6$ this shift saturates and the IX PL energy asymptotically approaches that of the indirect bandgap in bilayer WSe$_2$. We theoretically interpret this observation as the strong variation of the conduction band $K$ valley for $x>0.6$, with IX PL arising from the $K-K$ transition, while for $x<0.6$, the bandstructure hybridization becomes prevalent leading to the dominating momentum-indirect $K$-$Q$ transition. This bandstructure hybridization is accompanied with strong modification of IX PL dynamics and nonlinear exciton properties. Our work provides foundation for bandstructure engineering in van der Waals heterostructures highlighting the importance of hybridization effects and opening a way to devices with accurately tailored electronic properties.}

Van der Waals stacking allows unprecedented possibilities in combining different atomically thin materials within a single device as well as tuning of their properties, by selecting different monolayer materials or by varying their relative twist angle.  Thus, van der Waals (vdW) heterostructures made from vertically stacked atomically thin transition metal dichalcogenides (TMDs) present a powerful platform for development of novel devices with widely tuneable properties\cite{Geim2013,novoselov20162d,tang2021tuning} as well as for gaining new insights in the physics of excitons \textcolor{black}
{\cite{Kang2013,Gong2013,huang2022excitons,schmitt2022formation,li2021continuous}} and emergent excitonic and quantum phenomena in moiré superlattices \textcolor{black}{\cite{alexeev2019resonantly,wilson2021excitons,tran2019evidence,jones2021visualizing, Wilson2017,baek2020highly,tang2020simulation,xu2022tunable,shimazaki2020strongly}}. Most of the structures studied so far (such as for example stacked monolayers of MoSe$_2$ and WSe$_2$) were so-called type-II semiconducting heterostructures\cite{rivera2015observation}, exhibiting the maximum of the valence and the minimum of the conduction bands in the adjacent layers. A prominent feature in type-II TMD heterostructures is the formation of interlayer excitons (IX) with a hole and electron confined in different layers. The IX PL, although observed most clearly at cryogenic temperatures, has proven to be a sensitive probe of the novel physics in semiconducting van der Waals heterostructures\cite{rivera2018interlayer,jiang2021interlayer}, and will be utilised in our work to monitor the properties of the studied TMD heterobilayers. 

As well as the use of heterostructures, another very powerful approach for accurate tailoring of the properties of semiconductor devices is the use of alloyed semiconductors. This approach has been successfully applied in numerous applications, from lasers to few-nm-sized transistors using traditional III-V, II-VI and group IV semiconductors\cite{adachi2009properties}. This approach allows gradual tuning of bandgaps as well as the whole bandstructure, allowing to control carrier confinement and thus influence their transport and optical properties. Numerous semiconducting alloys of layered materials have been demonstrated in the monolayer form achieved both by direct synthesis and exfoliation from bulk\cite{xie2015two}, including Mo$_x$W$_{1-x}$Se$_2$ used in this work, where alloying was used to control the valley polarization properties\cite{Wang2015a, ye2017exciton}. More recently, alloying was successfully employed to tune the energy of interlayer excitons\cite{li2020wavelength} in WS$_{2(1-x)}$Se$_{2x}$/WSe$_2$, and a direct-indirect band transition in Mo$_{1-x}$W$_{x}$S$_2$/MoSe$_2$ heterostructures has been predicted theoretically \cite{zi2019reversible}. 

\begin{figure*}
	\includegraphics[width=1\textwidth]{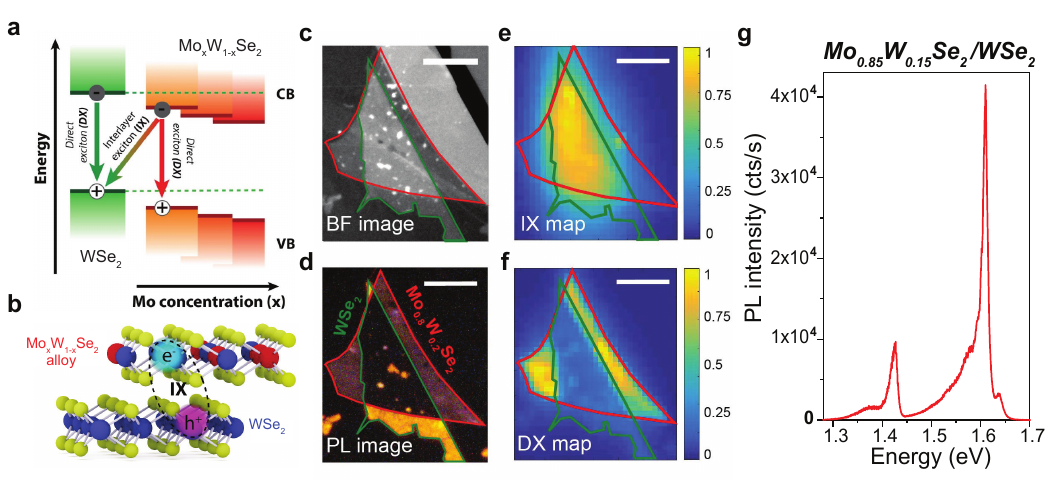}
	\caption{\large{\bf Van der Waals heterostructures based on TMD alloy monolayers.} \textbf{a,} Energy level diagram for a type-II heterojunction formed by WSe$_2$ and Mo$_{x}$W$_{1-x}$Se$_2$ monolayers. Direct excitons (DX) form in each layer, while interlayer excitons (IX) are created between electrons in the conduction band (CB) of the alloy and holes in the valence band (VB) of WSe$_2$. The IX energy is tuned by changing the Mo composition $x$ in the alloy. \textbf{b,} Schematic of the alloy HBL showing the IX electric dipole moment orientated perpendicular to the plane of the monolayers. \textbf{c,} Bright-field (BF) microscope image of the HBL. The red line shows the Mo$_{0.85}$W$_{0.15}$Se$_2$ flake while the green line shows the WSe$_2$ flake. \textbf{d,} PL image of the HBL taken at room temperature. PL quenching in the HBL region indicates efficient charge transfer between the monolayers. \textbf{e,} Low temperature ($T=10$ K) PL map of the HBL in the spectral region of IX PL (1.3-1.45 eV). Bright IX PL is detected from the HBL only. \textbf{f,} Low temperature (10 K) PL map of the HBL in the spectral region of DX PL (1.45-1.7 eV). Scale bars for (\textbf{c, d, e, f}) 10 $\mu$m. \textbf{g,} Low temperature (10K) spectrum measured for a Mo$_{0.85}$W$_{0.15}$Se$_2$ HBL. The dashed vertical line separates the two energy ranges for the DX and IX PL used for the PL maps in \textbf{e} and \textbf{f}.}\label{fig1}
\end{figure*}

In our work we study Mo$_x$W$_{1-x}$Se$_2$/WSe$_2$ and Mo$_x$W$_{1-x}$Se$_2$/MoSe$_2$ heterostructures built from stacked TMD monolayers in a wide range of composition $x$ (\textbf{Figure \ref{fig1}a}). In Mo$_x$W$_{1-x}$Se$_2$/WSe$_2$ we demonstrate tuning of $\approx$130 meV of the IX photoluminescence (PL) peak energy as a function of $x$ displaying two types of behaviour: a  fast tuning by $\approx 100$ meV for $x>0.6$ followed by an asymptotic dependence for $x<0.6$, where the IX PL energy asymptotically approaches that of the indirect bandgap in bilayer WSe$_2$. A similar but less pronounced tuning and saturation at the indirect bandgap of bilayer MoSe$_2$ is observed for Mo$_x$W$_{1-x}$Se$_2$/MoSe$_2$. In Mo$_x$W$_{1-x}$Se$_2$/WSe$_2$, concurrent with the transition between the two types of behaviour, we also observe strong changes in the IX PL dynamics and intensity: the PL lifetime shortens and intensity strongly decreases for $x<0.6$. The strong changes are also observed in the power dependence of IX PL, where a typical blue-shift of the IX PL peak \cite{rivera2015observation} is strongly suppressed for $x<0.6$. Our theoretical analysis shows that the observed changes in the behaviour occur due to the bandstructure hybridization, leading to the shift of the conduction band minimum in the heterobilayer from $K$ to $Q$ valley for $x\approx 0.6$. The observed behaviour thus signifies the transition from the $K-K$ configuration of the valence to conduction band transition for $x>0.6$ to the $K-Q$ configuration for $x<0.6$.


A set of 15 samples was studied, including heterobilayer (HBLs) either fully encapsulated in hBN flakes or alternatively placed on a relatively thick (of the order of 50 nm) hBN. The TMD HBLs were fabricated from monolayers mechanically exfoliated from bulk crystals. The alloy bulk crystals were provided by HQ Graphene, where the material composition was accurately measured in each bulk sample using energy dispersive X-ray (EDX) microanalysis.  The majority of HBLs were made with the prominent crystal edges aligned within 0.5° accuracy, constituting structures with either near 0° or 60° degree crystal axis rotation. A schematic of a generic Mo$_x$W$_{1-x}$Se$_2$/WSe$_2$ HBL is displayed in \textbf{Figure \ref{fig1}b}, also showing IX formation with a hole localized in WSe$_2$ and the electron in the alloy.

A bright field (BF) microscope image of a Mo$_{0.85}$W$_{0.15}$Se$_2$/WSe$_2$ heterostructure is shown in \textbf{Figure \ref{fig1}c}. The alloy monolayer area is highlighted with red lines, while the edges of the WSe$_2$ flake are shown in green. \textbf{Figure \ref{fig1}d} shows a PL image of this sample measured at room temperature  showing in yellow bright emission from the uncoupled monolayer WSe$_2$ and in purple dimmer emission from the uncoupled monolayer Mo$_{0.85}$W$_{0.15}$Se$_2$. Notably lower PL intensity is observed in the heterobilayer region where the two monolayers overlap. This arises from the quenching of the PL of the intralayer exciton (referred to below as DX) due to the efficient charge transfer between the layers. This shows efficient electronic coupling between the TMD monolayers, and confirms the formation of a type-II heterostructure\cite{Alexeev2017,Hong2014,Rivera2015}.

\textbf{Figure \ref{fig1}e,f} show PL maps of the same HBL acquired at a cryogenic temperature T = 10 K. The maps show integrated PL intensity in the spectral ranges 1.3-1.45 eV in Fig. \ref{fig1}e and 1.45-1.7 eV in Fig. \ref{fig1}f. These ranges are selected based on the low temperature PL spectra, with a typical example measured in the HBL area shown in \textbf{Figure \ref{fig1}g}. In addition to a strong peak around 1.6 eV corresponding to the intralayer exciton (DX) in Mo$_{0.85}$W$_{0.15}$Se$_2$ and some lower intensity features arising from WSe$_2$, additional peaks are observed at low energy (labelled IX). These peaks correspond to the interlayer exciton PL and are only observed within the HBL area.

\begin{figure}
\includegraphics[width=1\textwidth]{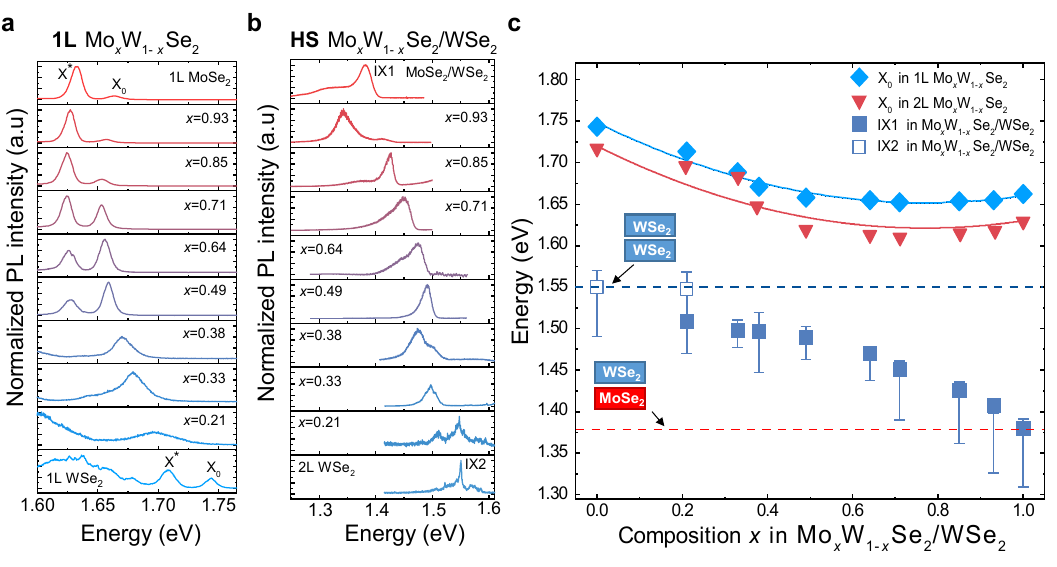}
\caption{\large {\bf Optical spectra of intralayer and interlayer excitons in alloy monolayers and heterostructures.} \textbf{a,} Normalized PL spectra measured at $T=10$K of Mo$_x$W$_{1-x}$Se$_2$ alloy monolayers with different Mo concentrations ($x$), showing the neutral exciton X$_0$ and the trion X* peaks. \textbf{b,} Normalized PL spectra of the Mo$_x$W$_{1-x}$Se$_2$/WSe$_2$ HBLs measured at 10K for different $x$. IX1 and IX2 label the interlayer exciton PL and the PL peak at the indirect bandgap in WSe$_2$ bilayers, respectively. \textbf{c,} Solid (open) squares show spectral positions of IX1 (IX2) PL peaks in the alloy/WSe$_2$ HBLs. See text for discussion of the $x=$0.21 HBL. Error bars show the spread in energy of multiple IX transition peaks measured in several HBL samples (see text for further details). Diamonds (triangles) show X$_0$ (A-exciton) peak position in alloy monolayers (bilayers) measured using reflectance contrast. Solid light blue and red lines are parabolic fits of the A-exciton peak energy in alloy monolayers and bilayers, respectively. The dashed red and blue horizontal lines correspond to the IX1 PL peak in MoSe$_2$/WSe$_2$ HBLs and the IX2 PL peak in WSe$_2$ homobilayers, respectively. }\label{fig2}
\end{figure}

Before describing the properties of HBLs comprising alloy monolayers, we briefly report optical properties of monolayer and bilayer alloy TMDs themselves. \textbf{Figure \ref{fig2}a} shows PL spectra for Mo$_x$W$_{1-x}$Se$_2$ alloy monolayers with different composition $x$. For $0.49 \leq x \leq 1$, the PL spectra closely resemble that of MoSe$_2$ with two narrow peaks attributed to neutral exciton, X$_0$ (the spectral feature of A-exciton) and the corresponding charged exciton or trion, X*. For $x\leq$ 0.38, the PL peaks become much broader, the X* peak becomes much less pronounced, and more features appear at lower energies. This behaviour is observed for concentrations for which the lowest neutral exciton state in the alloy transforms from optically bright as in MoSe$_2$ to optically dark as in WSe$_2$, as previously predicted \textcolor{black}{\cite{Wang2015a,kopaczek2021experimental}}. We have further studied reflectance contrast (RC) spectra from monolayers which allows detection of high energy states, for example B-excitons (see Supplementary Note S1). The A-exciton energy appears almost constant at 1.650 eV for 0.49 $\leq x \leq$ 1 with a weak increase towards $x=1$. For $x<0.49$, the energy of A peak increases eventually reaching 1.742 eV for monolayer WSe$_2$. The observed trend can be fitted by the quadratic Vegard's law (\textcolor{black}{Supplementary Note S1, Fig. \ref{fig2}c}), adopted for example for alloys of III-V semiconductors\cite{denton1991vegard}, revealing a bowing parameter b=0.16, in close agreement with previous reports\cite{Tongay2014,Zhang2014}. 
\textcolor{black}{Fig.S2 of the Supplementary Information} shows that B-exciton energy increases more sharply with decreasing $x$ from 1.877 to 2.185 eV signifying a strong increase in the valence band spin-orbit splitting as the A-B exciton splitting changes from $\approx 230$ to $\approx 340$ meV.

Of high relevance to HBL studies presented in this work we have also measured low T PL and RC for exfoliated Mo$_x$W$_{1-x}$Se$_2$ bilayers with different concentration of $x$ (see Supplementary Note S2). The RC spectra show a similar behaviour of A and B exciton energies compared to monolayers. In Fig. \ref{fig2}c we show a comparison between the A exciton peak positions in monolayers (diamonds) and bilayers (triangles), where similar parabolic trends are observed. However, a systematic red-shift up to 40 meV of the peak positions with respect to the monolayers with the same x occurs\cite{molas2017optical}. The PL data shows that the momentum-indirect transition is always dominant in bilayers (Supplementary Note S2), with its energy tuned by $>$ 100 meV by changing the Mo composition of the bilayer alloy Mo$_x$W$_{1-x}$Se$_2$.

We now discuss the tuning of the interlayer exciton energy in Mo$_x$W$_{1-x}$Se$_2$/WSe$_2$ HBLs as a function of $x$. \textbf{Figure \ref{fig2}b} shows normalized PL spectra measured for such structures at T = 10 K in the energy range where IX PL is expected, as identified in Fig. \ref{fig1}. Similarly to Fig. \ref{fig1}g, features corresponding to IX PL can be readily identified in the spectra of all structures with $x\geq$0.21. Following the approach depicted in Fig. \ref{fig1}, we measured PL maps at low temperature on all the structures to ensure that these features could be observed only in the regions where the two TMDs overlap (see Supplementary Note S3). For all HBLs with $x>$0.33, IX PL shows a spectrum with two or more peaks, similar to those reported for MoSe$_2$/WSe$_2$ heterostructures \cite{rivera2015observation,miller2017long,tran2019evidence,barre2022optical}. The energies of these PL peaks increase with decreasing $x$ (see a complete peak analysis of the IX spectra and a detailed discussion on the origin of the different spectral features in the Supplementary Note S4). The tuning is pronounced for $x$ up to 0.64 and exceeds 100 meV, but then slows down and is less than 50 meV for the whole range of $x<0.64$.

For $x=0.33$ these peaks practically merge into a single peak at $\approx$1.5 eV. For $x=0.21$, a peak at $\approx$1.51 eV provides a continuation of this trend, while new structured PL appears at higher energy at and above 1.54 eV. Similar high energy PL peaks are also observed around 1.54 eV and 1.57 eV in the natural WSe$_2$ bilayer \cite{Lindlau2018} (see the spectrum labelled 2L in the figure), while no discernible PL occurs in this sample at lower energies, where IX PL is observed in samples with $x>0$. Except for the IX feature  at $\approx$1.51 eV in the $x=0.21$ structure, the spectra for the structures with $x=0.21$ and $x=0$ are similar to the one reported for another WSe$_2$ bilayer in Fig.S3, measured on a different sample. 

Spectra similar to the ones reported in Fig. \ref{fig2}b were measured on 20 to 60 points across each HBL. Notably, the PL intensity averaged across the measured points of each HBL drops significantly for $x<0.6$ (see Supplementary Note S5). For some values of $x$, we also made several samples where similar PL measurements were carried out at multiple positions providing similar results.

The PL peak positions averaged over many measurements at different locations within each HBL are presented in Fig.\ref{fig2}c. Here the square symbols show the high energy peak position of the IX1 PL (E$_{IX}$), ascribed in MoSe$_2$/WSe$_2$ HBLs to the momentum-direct K-K transition \cite{miller2017long,tran2019evidence,barre2022optical}. The vertical bars show the spectral extension of the PL intensity within a standard deviation from the average intensity, covering all the features observed in the spectra near the IX energy. Overall, we observe that E$_{IX}$ changes with decreasing x by 130 meV from 1.38 eV for MoSe$_2$/WSe$_2$ to 1.51 eV for Mo$_{0.21}$W$_{0.79}$Se$_2$/WSe$_2$. This is in contrast to a weak dependence of the Mo$_x$W$_{1-x}$Se$_2$ bandgap (as can be deduced from a weak dependence of the X$_0$ exciton energy presented in the same plot) for $0.5 < x \leq 1$. Thus, in this range of x the conduction band offset of Mo$_x$W$_{1-x}$Se$_2$ with respect to that of WSe$_2$ evidently experiences a much faster change, with a shift by more than 100 meV, as can be deduced from the increase of E$_{IX}$ (assuming weak dependencies of the IX and DX exciton binding energies on $x$ \cite{ovesen2019interlayer}). 
It is seen from Fig.\ref{fig2}c that the dependence of E$_{IX}$ on $x$ is nonlinear and asymptotically approaches 1.51 eV, close to the energy of the exciton transition at the indirect bandgap in the homobilayer of WSe$_2$ marked IX2 in Fig.\ref{fig2}b  and shown with open squares in Fig.\ref{fig2}c.
 
We also fabricated several Mo$_x$W$_{1-x}$Se$_2$/MoSe$_2$ HBLs, where a more complicated behavior is observed for $0.33 < x \leq 1$, with less distinct PL peaks in the range where IX PL is expected (see Supplementary Note S6). However, with increasing $x$, the main feature of the low energy PL response asymptotically approaches the energy corresponding to PL at the indirect bandgap in homobilayer MoSe$_2$, similarly to the Mo$_x$W$_{1-x}$Se$_2$/WSe$_2$ HBLs.

\begin{figure}
	\includegraphics[width=0.8\textwidth]{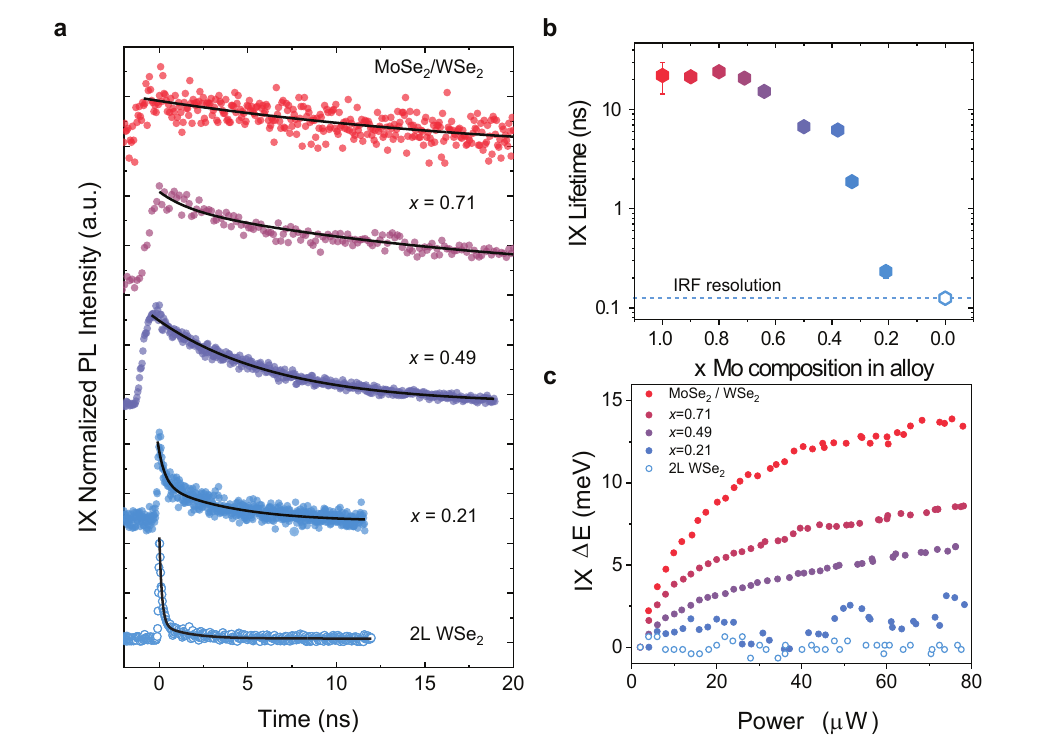}
	\caption{
		\large {\bf  Time-resolved and power-dependent PL of interlayer excitons in Mo$_x$W$_{1-x}$Se$_2$/WSe$_2$  heterostructures} 	
\textbf{a,} Time-resolved PL traces of IX in Mo$_x$W$_{1-x}$Se$_2$/WSe$_2$ HBLs with different $x$ measured at $T = 10$ K. Fitting with a single ($x=1, 0.49$) or double ($x=0.71, 0.21, 0$) exponential decay are shown with black lines. 
\textbf{b,} IX PL decay times as a function of $x$ for Mo$_x$W$_{1-x}$Se$_2$/WSe$_2$ HBLs extracted from the data in \textbf{a} . The dashed line indicates the temporal resolution limit given by the instrument response function (IRF) of the experimental setup. 
	\textbf{c,} Energy shift of the IX peak in Mo$_x$W$_{1-x}$Se$_2$/WSe$_2$ HBLs with different $x$ as a function of the excitation power.} \label{fig3}
\end{figure}

In order to understand the nature of the low energy PL peaks in the studied HBLs and to gain further insight in the properties of the IX as a function of the alloy composition, we carried out time-resolved and power-dependent PL measurements. The transient PL decay signals measured for Mo$_x$W$_{1-x}$Se$_2$/WSe$_2$ HBLs are shown in \textbf{Figure \ref{fig3}a}. Each transient PL decay curve is measured at the IX1 PL peak shown in Fig. \ref{fig2}c. For Mo$_{0.21}$W$_{0.79}$Se$_2$/WSe$_2$ structure, we measure at the energy of 1.51 eV corresponding to the energy of IX1 (solid square in Fig. \ref{fig2}c). For the WSe$_2$ bilayer the measurement is carried out at the PL peak around 1.55 eV (IX2, open squares in Fig. \ref{fig2}c).

It can be observed that the PL decay of the IX shortens progressively, decreasing from tens of nanoseconds in MoSe$_2$/WSe$_2$ down to hundreds of picoseconds in the Mo$_{0.21}$W$_{0.79}$Se$_2$/WSe$_2$ HBL and bilayer WSe$_2$. The lifetime values extracted from exponential fits of the time-resolved PL curves are summarized in Fig. \ref{fig3}b as a function of the alloy composition. As the alloys gradually become more chemically similar to WSe$_2$, the PL lifetime decreases by more than two orders of magnitude, with the most abrupt change occurring for $x<0.4$. The fast decay times extracted for the WSe$_2$ bilayer are limited by the temporal resolution of the set-up (shown with a dashed line in Fig. \ref{fig3}b) and are in agreement with the 25 ps lifetime previously reported for WSe$_2$ bilayers\cite{Wang2014a}. Such a strong reduction of the PL lifetime may arise from two related effects. Firstly, a similar effect has been observed in a gated bilayer WSe$_2$ \cite{Wang2018} , where the application of a vertical electric field induces an increase of the interlayer exciton lifetime, directly related to the enhanced spatial confinement of the carriers in the adjacent layers. In our case, gradual tuning of the band structure towards that of a WSe$_2$ bilayer should result in an increased spatial overlap of the electron and hole wavefunctions, leading to interlayer excitons with shorter lifetimes. On the other hand, as we discuss below, a gradual transition to the indirect bandgap semiconductor with decreasing $x$ should lead to strong nonradiative decay \cite{wang2014exciton}, as additionally evidenced by the reduction of the overall PL emission in HBLs with small $x$ (see Supplementary Figure S9).

Further confirmation of this interpretation is obtained from the power dependence of the spectral position of the IX PL, which can be used to assess the efficiency of the exciton-exciton interaction between IXs. The IXs possess permanent dipole moments, aligned along the direction normal to the plane of the device, resulting in enhanced repulsive dipolar exciton-exciton interactions\cite{rivera2015observation}. This gives rise to a blue-shift of the IX PL peak energy at sufficiently high excitation density, i.e. for high IX densities. Above a certain power, both the blue-shift and the PL intensity saturate, due to the exciton-exciton annihilation processes\cite{sun2014observation}. Fig. \ref{fig3}c shows experimentally measured blue-shift as a function of the power of a continuous-wave laser in several HBLs with different $x$. We observe a maximum shift of 14 meV in a MoSe$_2$/WSe$_2$ structures at the excitation power of 80 $\mu W$ in agreement with previously reported values \cite{rivera2015observation}. The shift progressively decreases for smaller $x$ and as the heterobilayer becomes more similar to a homobilayer, the blue-shift gradually disappears. At the same time, the power-dependences of PL intensities, showing saturation for high Mo concentrations, become nearly linear as $x$ decreases (see Supplementary Note S8). 

We believe that these observations support the conclusions derived from the PL dynamics. For decreasing $x$, the electron and hole within the IX-like particle become less localized in the individual layers. This will lead to a smaller dipole moment and thus weaker exciton-exciton interaction and a reduced blue-shift. At the same time, and possibly with a stronger influence on the observed behaviour, as the bandgap becomes indirect at small Mo concentrations, the fast non-radiative decay prevents build-up of the exciton population thus suppressing the exciton-exciton interactions and the corresponding blue-shift.

\begin{figure}
\includegraphics[width=1\textwidth]{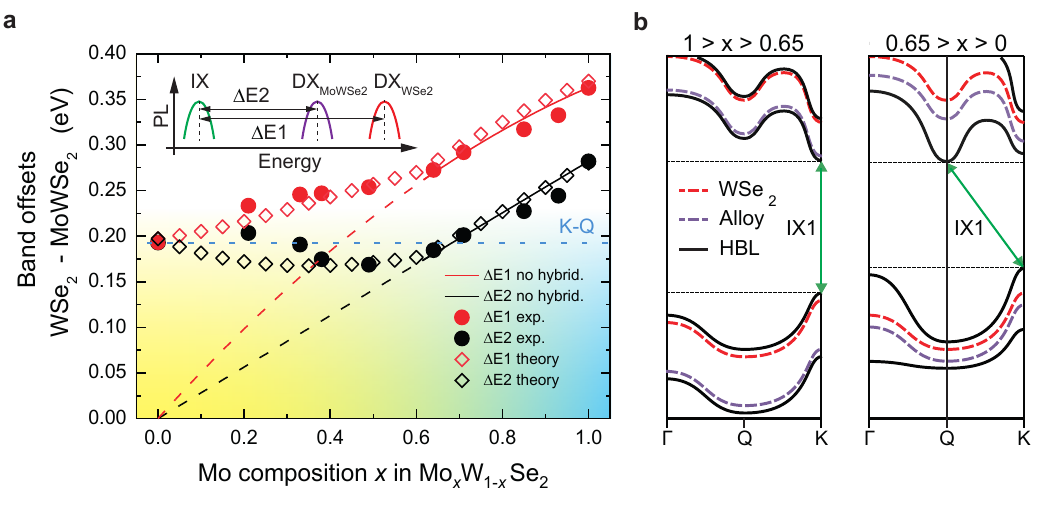}
\caption{\large {\bf Band offsets of the Mo$_x$W$_{1-x}$Se$_2$/WSe$_2$ HBLs.} \textbf{a,} Band offsets of Mo$_x$W$_{1-x}$Se$_2$/WSe$_2$ HBLs as functions of Mo composition x in the alloy layers. The band offsets $\Delta$E1 and $\Delta$E2 are extracted as shown in the diagram in \textbf{a} from the experimentally measured PL peaks of IX in the HBLs and direct excitons DX$_{MoWSe_2}$ in alloys and DX$_{WSe_2}$ in WSe$_2$ measured in isolated monolayers. Experimentally measured $\Delta$E1 and $\Delta$E2 are displayed as filled red and black circles, respectively. Red (black) solid and dashed lines show theoretical values for the conduction (valence) band offsets, obtained assuming a direct HBL bandgap at the $K$ point, and neglecting interlayer band hybridization. The experimental energy offsets deviate significantly from the predicted trend for $x\lesssim 0.6$, suggesting strong interlayer band hybridization for that range of Mo concentrations. Empty diamonds show the theoretically predicted trends for $\Delta E1$ and $\Delta E2$ when interlayer hybridization at the $K$, $Q$ and $\Gamma$ points of the Brillouin zone is considered, assuming  an interlayer angle of $60^\circ$, showing excellent agreement with the experimental data for all $x$ values. \textbf{b,} Sketch of the $x$-dependent HBL band structure evolution, according to the interlayer band hybridization model. Dashed (solid) lines indicate the monolayer (HBL) band energies. Whereas for $x>0.65$ the HBL remains direct due to weak hybridization at the $K$ point, strong hybridization of conduction states shifts the conduction band edge to the $Q$ point for $x<0.65$, resulting in an indirect $K-Q$ bandgap, and a finite momentum for the IX.}\label{fig4}
\end{figure}

We now further analyse the PL data and interpret the asymptotic behaviour of the IX energy as a function of $x$. The conduction- and valence band energy offsets between the WSe$_2$ and alloy layers forming a given HBL, $\Delta E1$ and $\Delta E2$, can be computed by subtracting the HBL bandgap from the intralayer bandgaps of the WSe$_2$ and alloy layers, respectively (see the diagram in Fig. \ref{fig4}a). To this end, we estimate the HBL bandgap as the IX energy, and the intralayer gaps by the corresponding direct exciton energies, as obtained from PL measurements reported in Fig. \ref{fig2}a and b. In this approximation, we are neglecting the exciton binding energies and their $x$ dependences\cite{ovesen2019interlayer}, thus incurring an error of the order of tens of meV. The experimental $x$-dependent energy offsets $\Delta E1$ and $\Delta E2$ are reported in Fig. \ref{fig4}a, where two distinct trends can be observed for both curves, one for $0\le x \lesssim 0.6$ and another for $0.6\lesssim x \le 1$.

The band offset trends measured for $0.6\lesssim x \le 1$ can be explained by simple interpolation of the conduction and valence band edge energies of the alloy, assigning a linear trend to the latter, and a quadratic trend to the former with the same bowing parameter measured for the direct A exciton in the alloy\cite{denton1991vegard}. Such interpolation neglects the effects of hybridization between the WSe${}_2$ and alloy layers, and assumes that, as is the case in MoSe$_2$/WSe$_2$ ($x=1$)\cite{Wilson2017,barre2022optical}, the HBL bandgap remains at the $K$ point for all values of $x$. Full details of this approximation can be found in Supplementary Note S7. The resulting interpolations, shown with solid and dashed lines in Fig. \ref{fig4}a, match the experimental trends for $\Delta E1$ and $\Delta E2$ only for $0.6\lesssim x \le 1$, indicating that interlayer hybridization in the HBL is weak for this range of Mo compositions, and that, indeed, the HBL bandgap remains at the $K$ point. However, continuing the interpolation down to $x \lesssim 0.6$ we find a clear and dramatic deviation from the experimental values, where the predicted offsets (dashed lines in Fig. \ref{fig4}a) vanish for $x=0$, while the experimental values (circles in Fig.\ref{fig4}a) converge at a saturation value around 200 meV. This deviation can be explained only if we consider major modifications to the HBL band structure for $0\le x \lesssim 0.6$, such as a transition to an indirect bandgap for the HBL, caused by strong interlayer hybridization at the $\Gamma$ and $Q$ points of the Brillouin zone\cite{kravtsov2021spin}.

To explore this possibility, we have formulated simple Hamiltonians for interlayer hybridization of the WSe$_2$ and alloy bands in the HBL, at the $K$ (conduction and valence), $\Gamma$ (valence) and $Q$ (conduction) points, with parameters based on density functional theory calculations extracted from Ref.\ \cite{magorrian2021multifaceted}. It is worthwhile mentioning that these models contain no free parameters. Further model details can be found in Supplementary Note S7. Importantly, symmetry considerations\cite{magorrian2021multifaceted} dictate that the model parameters differ for HBLs with interlayer angles close to $0^\circ$ and $60^\circ$. We have determined that the best match to the experimental data is obtained assuming HBLs with $60^\circ$ alignment (see Supplementary Fig.\ S14 in Supplementary Note S7), and limit our discussion to that case. The corresponding predictions for the band offsets $\Delta E1$ and $\Delta E2$, shown with red and black diamonds in Fig.\ \ref{fig4}a, give an excellent quantitative match to the experimental data throughout the full range of compositions, and importantly with the saturation value of approximately $200\,{\rm meV}$ at $x=0$. The model shows that the HBL conduction band edge migrates from the $K$ to the $Q$ point for $x<0.65$, whereas the HBL valence band edge remains at the $K$ point, as illustrated in Fig.\ \ref{fig4}b, consistent with recent theoretical reports for WSe$_2$ bilayers\cite{magorrian2021multifaceted}. This results in an indirect ($K-Q$) bandgap that changes the nature of the IX, thus shedding light on the change of trend of $\Delta E1$ and $\Delta E2$, the band offset saturation for $x=0$, and the suppression of the IX energy tuning with $x$ for Mo concentrations below $0.65$. Moreover, the quantitative agreement between the model and experimental data strongly suggests that the samples studied are approximately H-stacked HBLs, with an interlayer twist angle close to $60^\circ$. 

We now briefly comment on the simultaneous observation of IX1 peak and a high energy PL feature around 1.55 eV (similar to IX2 in bilayer WSe$_2$) in the HBL with $x=$0.21. While IX1 PL originates from the $K-Q$ transition, as suggested above, the following options for the origin of the 1.55 eV feature may be proposed. As can be deduced from the broad spectral features in Mo$_{0.21}$W$_{0.79}$Se$_2$/WSe$_2$ alloy mono- and bilayers, this material is highly disordered. This may lead to occurrence of W-rich clusters, which will locally behave as WSe$_2$ bilayers, thus giving rise to PL at the IX2 energy. Another explanation, assuming a more even distribution of Mo in the alloy is related to PL emission involving $\Gamma-Q$ transition, expected around 50 meV higher than $K-Q$, and activated in this alloy either due to the increased disorder or a relatively high hole concentration.

In summary, we have demonstrated significant tuning of IX energy in Mo$_x$W$_{1-x}$Se$_2$/WSe$_2$ and Mo$_x$W$_{1-x}$Se$_2$/MoSe$_2$ alloy heterostructures by changing the Mo composition $x$. We use PL measurements at 10 K, where we observe a shift of IX PL energy by 130 meV in Mo$_x$W$_{1-x}$Se$_2$/WSe$_2$ as a function of $x$, as well as an unusual asymptotic dependence of IX PL energy approaching the homobilayer configuration. We show that this behaviour stems from strong hybridization between conduction states of the WSe$_2$ and alloyed layers for $x<0.65$, when the interlayer conduction band offset falls below $\approx 250$ meV, leading to a $K-Q$ indirect bandgap for the heterostructure, and thus changing the nature of the IX. We probed the effects of such strong band structure modifications on IX also by means of time-resolved PL experiments and by measuring PL power dependences, supporting our conclusion about the transition to an indirect bandgap in the HBLs with $x<0.65$. We also find similar effects in Mo$_x$W$_{1-x}$Se$_2$/MoSe$_2$ HBLs. As background information, we also extract detailed energy dependences of A- and B- excitons, A* trions and IX in monolayer and bilayer Mo$_x$W$_{1-x}$Se$_2$ as a function of $x$.

This work demonstrates the potential of adopting TMD alloys in vdW heterobilayers to achieve continuous band structure tuning, and fine band hybridization engineering. Tuning the A-exciton energy and band-edge offsets using alloying provides an additional probing approach for lattice reconstruction phenomena, currently actively being discussed in theory and observed in microscopy experiments. 

\section{Methods}


 High-quality fully encapsulated HBL samples were fabricated using PMMA-assisted dry-peel transfer. To minimize contamination, heterostructures were fabricated using a remotely controlled micromanipulation setup placed inside an argon chamber with $<$0.1 p.p.m. O$_2$ and H$_2$O. The bulk crystals were mechanically exfoliated onto a 90 nm layer of PMMA coated on a silicon wafer. Monolayers were then identified via optical microscopy, as well as through luminescence imaging in the dark-field configuration.  Crystals that had adjacent straight edges at 0$^{\circ}$, 60$^{\circ}$ or 120$^{\circ}$ to one another (indicating one of the crystallographic axes) were then selected and picked up onto an hBN film (less than around 10 nm thickness) held by a PMMA membrane. During the transfer of the second TMD layer the edges were aligned to within 0.5$^{\circ}$ of the desired angle and finally transferred onto another hBN film (less or around 20 nm thickness) exfoliated onto an oxidized silicon wafer (70 nm SiO$_2$) to achieve full encapsulation. To prevent spontaneous rotation of the TMD layers and deterioration of MoSe$_2$ crystalline quality we avoided exposing the heterostructures to temperatures greater than 70$^{\circ}$C.


The photoluminescence images of the heterobilayer samples were acquired using a modified bright-field microscope (LV150N, Nikon) equipped with a colour camera (DS-Vi1, Nikon). The near-infrared emission from the white-light source was blocked with a 550-nm short-pass filter (FESH0550, Thorlabs), and a 600-nm long-pass filter (FELH0600, Thorlabs) was used to isolate the photoluminescence signal from the sample.

Spectrally-resolved photoluminescence and reflectance contrast measurements were performed using a custom-built micro-photoluminescence setup. For photoluminescence, the excitation light centred at 2.33 eV was generated by a diode-pumped solid-state laser (CW532-050, Roithner), whereas for reflectance contrast a stabilized tungsten-halogen white-light source (SLS201L, Thorlabs) was used. The excitation light was focused onto the sample using a 50x objective lens (M Plan Apo 50X, Mitutoyo). The photoluminescence and reflectance contrast signals collected in the backwards direction were detected by a 0.5-m spectrometer (SP-2-500i, Princeton Instruments) with a nitrogen-cooled charge-coupled device camera (PyLoN:100BR, Princeton Instruments). The photoluminescence signal was isolated using a 550-nm short-pass filter (FELH0550, Thorlabs). The reflectance contrast spectra were derived by comparing the spectra of white light reflected from the sample and the substrate as RC($\lambda$) = (R($\lambda$) - R0($\lambda$))/(R0($\lambda$)), where R (R0) is the intensity of light reflected by the sample (substrate). The room-temperature measurements were performed in ambient conditions. The low-temperature measurements were carried out using a continuous-flow liquid helium cryostat, in which the sample was placed on a cold finger with a base temperature of 10 K. 

To acquire the PL decay of the IX in the entire set of samples we collected the emitted photons with an avalanche diode photodetector (APD) (ID100-MMF50), with a timing resolution of $ \sim $40 ps, and a photon counting card (SPC-130). A 638 nm pulsed diode laser (PicoQuant LDH) is used as 
excitation source, with 80 MHz repetition rate, which results in a instrument response function (IRF) with $ \sim $150 ps FWHM.

\section{Acknowledgements}

AC, LC, RP, KE, AIT acknowledge financial support of the European Commission H2020-MSCA-ITN project under grant agreements 676108. AG, CL, DJG, EA and AT acknowledge financial support of the European Graphene Flagship Projects under grant agreements 785219 and 881603 and EPSRC grants EP/V006975/1,  EP/V026496/1, EP/V034804/1 and EP/S030751/1. VF and KE acknowledge financial support of the European Graphene Flagship Project under grant agreement 881603. AG acknowledges support by the European Union project ENOSIS H2020-MSCA-IF-2020-101029644. K.W. and T.T. acknowledge support from the JSPS KAKENHI (Grant Numbers 20H00354, 21H05233 and 23H02052) and World Premier International Research Centre Initiative (WPI), MEXT, Japan. DART acknowledges funding from PAPIIT-DGAPA-UNAM grant IA106523.

\section{Author contributions}

AC, RP, LH, AK and CL fabricated heterostructure samples. KW and TT synthesized the high quality hBN. AC, AG, CL, DJG, LS and EA carried out optical spectroscopy measurements. AC, AG and CL analyzed the data with contribution from DJG, LS, EA, AIT, DART. DART and VF developed theory. AG, CL, AIT and DART wrote the manuscript with contribution from all co-authors. AIT, KE, KSN managed various experimental aspects of the project. AIT conceived and supervised the project.

\bibliography{sn-bibliography}

\begin{thebibliography}{44}%
\makeatletter
\providecommand \@ifxundefined [1]{%
 \@ifx{#1\undefined}
}%
\providecommand \@ifnum [1]{%
 \ifnum #1\expandafter \@firstoftwo
 \else \expandafter \@secondoftwo
 \fi
}%
\providecommand \@ifx [1]{%
 \ifx #1\expandafter \@firstoftwo
 \else \expandafter \@secondoftwo
 \fi
}%
\providecommand \natexlab [1]{#1}%
\providecommand \enquote  [1]{``#1''}%
\providecommand \bibnamefont  [1]{#1}%
\providecommand \bibfnamefont [1]{#1}%
\providecommand \citenamefont [1]{#1}%
\providecommand \href@noop [0]{\@secondoftwo}%
\providecommand \href [0]{\begingroup \@sanitize@url \@href}%
\providecommand \@href[1]{\@@startlink{#1}\@@href}%
\providecommand \@@href[1]{\endgroup#1\@@endlink}%
\providecommand \@sanitize@url [0]{\catcode `\\12\catcode `\$12\catcode
  `\&12\catcode `\#12\catcode `\^12\catcode `\_12\catcode `\%12\relax}%
\providecommand \@@startlink[1]{}%
\providecommand \@@endlink[0]{}%
\providecommand \url  [0]{\begingroup\@sanitize@url \@url }%
\providecommand \@url [1]{\endgroup\@href {#1}{\urlprefix }}%
\providecommand \urlprefix  [0]{URL }%
\providecommand \Eprint [0]{\href }%
\providecommand \doibase [0]{https://doi.org/}%
\providecommand \selectlanguage [0]{\@gobble}%
\providecommand \bibinfo  [0]{\@secondoftwo}%
\providecommand \bibfield  [0]{\@secondoftwo}%
\providecommand \translation [1]{[#1]}%
\providecommand \BibitemOpen [0]{}%
\providecommand \bibitemStop [0]{}%
\providecommand \bibitemNoStop [0]{.\EOS\space}%
\providecommand \EOS [0]{\spacefactor3000\relax}%
\providecommand \BibitemShut  [1]{\csname bibitem#1\endcsname}%
\let\auto@bib@innerbib\@empty
\bibitem [{\citenamefont {Geim}\ and\ \citenamefont {et~al.}(2013)}]{Geim2013}%
  \BibitemOpen
  \bibfield  {author} {\bibinfo {author} {\bibfnamefont {A.~K.}\ \bibnamefont
  {Geim}}\ and\ \bibinfo {author} {\bibnamefont {et~al.}},\ }\bibfield  {title}
  {\bibinfo {title} {{Van der Waals heterostructures}},\ }\href@noop {}
  {\bibfield  {journal} {\bibinfo  {journal} {Nature}\ }\textbf {\bibinfo
  {volume} {499}},\ \bibinfo {pages} {419} (\bibinfo {year}
  {2013})}\BibitemShut {NoStop}%
\bibitem [{\citenamefont {Novoselov}\ \emph {et~al.}(2016)\citenamefont
  {Novoselov}, \citenamefont {Mishchenko}, \citenamefont {Carvalho},\ and\
  \citenamefont {Neto}}]{novoselov20162d}%
  \BibitemOpen
  \bibfield  {author} {\bibinfo {author} {\bibfnamefont {K.}~\bibnamefont
  {Novoselov}}, \bibinfo {author} {\bibfnamefont {A.}~\bibnamefont
  {Mishchenko}}, \bibinfo {author} {\bibfnamefont {A.}~\bibnamefont
  {Carvalho}},\ and\ \bibinfo {author} {\bibfnamefont {A.~C.}\ \bibnamefont
  {Neto}},\ }\bibfield  {title} {\bibinfo {title} {2d materials and van der
  waals heterostructures},\ }\href@noop {} {\bibfield  {journal} {\bibinfo
  {journal} {Science}\ }\textbf {\bibinfo {volume} {353}},\ \bibinfo {pages}
  {aac9439} (\bibinfo {year} {2016})}\BibitemShut {NoStop}%
\bibitem [{\citenamefont {Tang}\ \emph {et~al.}(2021)\citenamefont {Tang},
  \citenamefont {Gu}, \citenamefont {Liu}, \citenamefont {Watanabe},
  \citenamefont {Taniguchi}, \citenamefont {Hone}, \citenamefont {Mak},\ and\
  \citenamefont {Shan}}]{tang2021tuning}%
  \BibitemOpen
  \bibfield  {author} {\bibinfo {author} {\bibfnamefont {Y.}~\bibnamefont
  {Tang}}, \bibinfo {author} {\bibfnamefont {J.}~\bibnamefont {Gu}}, \bibinfo
  {author} {\bibfnamefont {S.}~\bibnamefont {Liu}}, \bibinfo {author}
  {\bibfnamefont {K.}~\bibnamefont {Watanabe}}, \bibinfo {author}
  {\bibfnamefont {T.}~\bibnamefont {Taniguchi}}, \bibinfo {author}
  {\bibfnamefont {J.}~\bibnamefont {Hone}}, \bibinfo {author} {\bibfnamefont
  {K.~F.}\ \bibnamefont {Mak}},\ and\ \bibinfo {author} {\bibfnamefont
  {J.}~\bibnamefont {Shan}},\ }\bibfield  {title} {\bibinfo {title} {Tuning
  layer-hybridized moir{\'e} excitons by the quantum-confined stark effect},\
  }\href@noop {} {\bibfield  {journal} {\bibinfo  {journal} {Nature
  Nanotechnology}\ }\textbf {\bibinfo {volume} {16}},\ \bibinfo {pages} {52}
  (\bibinfo {year} {2021})}\BibitemShut {NoStop}%
\bibitem [{\citenamefont {Kang}\ and\ \citenamefont {et~al.}(2013)}]{Kang2013}%
  \BibitemOpen
  \bibfield  {author} {\bibinfo {author} {\bibfnamefont {J.}~\bibnamefont
  {Kang}}\ and\ \bibinfo {author} {\bibnamefont {et~al.}},\ }\bibfield  {title}
  {\bibinfo {title} {{Band offsets and heterostructures of two-dimensional
  semiconductors}},\ }\href@noop {} {\bibfield  {journal} {\bibinfo  {journal}
  {Applied Physics Letters}\ }\textbf {\bibinfo {volume} {102}},\ \bibinfo
  {pages} {22} (\bibinfo {year} {2013})}\BibitemShut {NoStop}%
\bibitem [{\citenamefont {Gong}\ and\ \citenamefont {et~al.}(2013)}]{Gong2013}%
  \BibitemOpen
  \bibfield  {author} {\bibinfo {author} {\bibfnamefont {C.}~\bibnamefont
  {Gong}}\ and\ \bibinfo {author} {\bibnamefont {et~al.}},\ }\bibfield  {title}
  {\bibinfo {title} {{Band alignment of two-dimensional transition metal
  dichalcogenides: Application in tunnel field effect transistors}},\
  }\href@noop {} {\bibfield  {journal} {\bibinfo  {journal} {Applied Physical
  Letters}\ }\textbf {\bibinfo {volume} {103}},\ \bibinfo {pages} {1} (\bibinfo
  {year} {2013})}\BibitemShut {NoStop}%
\bibitem [{\citenamefont {Huang}\ \emph {et~al.}(2022)\citenamefont {Huang},
  \citenamefont {Choi}, \citenamefont {Shih},\ and\ \citenamefont
  {Li}}]{huang2022excitons}%
  \BibitemOpen
  \bibfield  {author} {\bibinfo {author} {\bibfnamefont {D.}~\bibnamefont
  {Huang}}, \bibinfo {author} {\bibfnamefont {J.}~\bibnamefont {Choi}},
  \bibinfo {author} {\bibfnamefont {C.-K.}\ \bibnamefont {Shih}},\ and\
  \bibinfo {author} {\bibfnamefont {X.}~\bibnamefont {Li}},\ }\bibfield
  {title} {\bibinfo {title} {Excitons in semiconductor moir{\'e}
  superlattices},\ }\href@noop {} {\bibfield  {journal} {\bibinfo  {journal}
  {Nature Nanotechnology}\ }\textbf {\bibinfo {volume} {17}},\ \bibinfo {pages}
  {227} (\bibinfo {year} {2022})}\BibitemShut {NoStop}%
\bibitem [{\citenamefont {Schmitt}\ \emph {et~al.}(2022)\citenamefont
  {Schmitt}, \citenamefont {Bange}, \citenamefont {Bennecke}, \citenamefont
  {AlMutairi}, \citenamefont {Meneghini}, \citenamefont {Watanabe},
  \citenamefont {Taniguchi}, \citenamefont {Steil}, \citenamefont {Luke},
  \citenamefont {Weitz} \emph {et~al.}}]{schmitt2022formation}%
  \BibitemOpen
  \bibfield  {author} {\bibinfo {author} {\bibfnamefont {D.}~\bibnamefont
  {Schmitt}}, \bibinfo {author} {\bibfnamefont {J.~P.}\ \bibnamefont {Bange}},
  \bibinfo {author} {\bibfnamefont {W.}~\bibnamefont {Bennecke}}, \bibinfo
  {author} {\bibfnamefont {A.}~\bibnamefont {AlMutairi}}, \bibinfo {author}
  {\bibfnamefont {G.}~\bibnamefont {Meneghini}}, \bibinfo {author}
  {\bibfnamefont {K.}~\bibnamefont {Watanabe}}, \bibinfo {author}
  {\bibfnamefont {T.}~\bibnamefont {Taniguchi}}, \bibinfo {author}
  {\bibfnamefont {D.}~\bibnamefont {Steil}}, \bibinfo {author} {\bibfnamefont
  {D.~R.}\ \bibnamefont {Luke}}, \bibinfo {author} {\bibfnamefont {R.~T.}\
  \bibnamefont {Weitz}}, \emph {et~al.},\ }\bibfield  {title} {\bibinfo {title}
  {Formation of moir{\'e} interlayer excitons in space and time},\ }\href@noop
  {} {\bibfield  {journal} {\bibinfo  {journal} {Nature}\ }\textbf {\bibinfo
  {volume} {608}},\ \bibinfo {pages} {499} (\bibinfo {year}
  {2022})}\BibitemShut {NoStop}%
\bibitem [{\citenamefont {Li}\ \emph {et~al.}(2021)\citenamefont {Li},
  \citenamefont {Jiang}, \citenamefont {Li}, \citenamefont {Zhang},
  \citenamefont {Kang}, \citenamefont {Zhu}, \citenamefont {Watanabe},
  \citenamefont {Taniguchi}, \citenamefont {Chowdhury}, \citenamefont {Fu}
  \emph {et~al.}}]{li2021continuous}%
  \BibitemOpen
  \bibfield  {author} {\bibinfo {author} {\bibfnamefont {T.}~\bibnamefont
  {Li}}, \bibinfo {author} {\bibfnamefont {S.}~\bibnamefont {Jiang}}, \bibinfo
  {author} {\bibfnamefont {L.}~\bibnamefont {Li}}, \bibinfo {author}
  {\bibfnamefont {Y.}~\bibnamefont {Zhang}}, \bibinfo {author} {\bibfnamefont
  {K.}~\bibnamefont {Kang}}, \bibinfo {author} {\bibfnamefont {J.}~\bibnamefont
  {Zhu}}, \bibinfo {author} {\bibfnamefont {K.}~\bibnamefont {Watanabe}},
  \bibinfo {author} {\bibfnamefont {T.}~\bibnamefont {Taniguchi}}, \bibinfo
  {author} {\bibfnamefont {D.}~\bibnamefont {Chowdhury}}, \bibinfo {author}
  {\bibfnamefont {L.}~\bibnamefont {Fu}}, \emph {et~al.},\ }\bibfield  {title}
  {\bibinfo {title} {Continuous mott transition in semiconductor moir{\'e}
  superlattices},\ }\href@noop {} {\bibfield  {journal} {\bibinfo  {journal}
  {Nature}\ }\textbf {\bibinfo {volume} {597}},\ \bibinfo {pages} {350}
  (\bibinfo {year} {2021})}\BibitemShut {NoStop}%
\bibitem [{\citenamefont {Alexeev}\ \emph {et~al.}(2019)\citenamefont
  {Alexeev}, \citenamefont {Ruiz-Tijerina}, \citenamefont {Danovich},
  \citenamefont {Hamer}, \citenamefont {Terry}, \citenamefont {Nayak},
  \citenamefont {Ahn}, \citenamefont {Pak}, \citenamefont {Lee}, \citenamefont
  {Sohn} \emph {et~al.}}]{alexeev2019resonantly}%
  \BibitemOpen
  \bibfield  {author} {\bibinfo {author} {\bibfnamefont {E.~M.}\ \bibnamefont
  {Alexeev}}, \bibinfo {author} {\bibfnamefont {D.~A.}\ \bibnamefont
  {Ruiz-Tijerina}}, \bibinfo {author} {\bibfnamefont {M.}~\bibnamefont
  {Danovich}}, \bibinfo {author} {\bibfnamefont {M.~J.}\ \bibnamefont {Hamer}},
  \bibinfo {author} {\bibfnamefont {D.~J.}\ \bibnamefont {Terry}}, \bibinfo
  {author} {\bibfnamefont {P.~K.}\ \bibnamefont {Nayak}}, \bibinfo {author}
  {\bibfnamefont {S.}~\bibnamefont {Ahn}}, \bibinfo {author} {\bibfnamefont
  {S.}~\bibnamefont {Pak}}, \bibinfo {author} {\bibfnamefont {J.}~\bibnamefont
  {Lee}}, \bibinfo {author} {\bibfnamefont {J.~I.}\ \bibnamefont {Sohn}}, \emph
  {et~al.},\ }\bibfield  {title} {\bibinfo {title} {Resonantly hybridized
  excitons in moir{\'e} superlattices in van der waals heterostructures},\
  }\href@noop {} {\bibfield  {journal} {\bibinfo  {journal} {Nature}\ }\textbf
  {\bibinfo {volume} {567}},\ \bibinfo {pages} {81} (\bibinfo {year}
  {2019})}\BibitemShut {NoStop}%
\bibitem [{\citenamefont {Wilson}\ \emph {et~al.}(2021)\citenamefont {Wilson},
  \citenamefont {Yao}, \citenamefont {Shan},\ and\ \citenamefont
  {Xu}}]{wilson2021excitons}%
  \BibitemOpen
  \bibfield  {author} {\bibinfo {author} {\bibfnamefont {N.~P.}\ \bibnamefont
  {Wilson}}, \bibinfo {author} {\bibfnamefont {W.}~\bibnamefont {Yao}},
  \bibinfo {author} {\bibfnamefont {J.}~\bibnamefont {Shan}},\ and\ \bibinfo
  {author} {\bibfnamefont {X.}~\bibnamefont {Xu}},\ }\bibfield  {title}
  {\bibinfo {title} {Excitons and emergent quantum phenomena in stacked 2d
  semiconductors},\ }\href@noop {} {\bibfield  {journal} {\bibinfo  {journal}
  {Nature}\ }\textbf {\bibinfo {volume} {599}},\ \bibinfo {pages} {383}
  (\bibinfo {year} {2021})}\BibitemShut {NoStop}%
\bibitem [{\citenamefont {Tran}\ \emph {et~al.}(2019)\citenamefont {Tran},
  \citenamefont {Moody}, \citenamefont {Wu}, \citenamefont {Lu}, \citenamefont
  {Choi}, \citenamefont {Kim}, \citenamefont {Rai}, \citenamefont {Sanchez},
  \citenamefont {Quan}, \citenamefont {Singh} \emph
  {et~al.}}]{tran2019evidence}%
  \BibitemOpen
  \bibfield  {author} {\bibinfo {author} {\bibfnamefont {K.}~\bibnamefont
  {Tran}}, \bibinfo {author} {\bibfnamefont {G.}~\bibnamefont {Moody}},
  \bibinfo {author} {\bibfnamefont {F.}~\bibnamefont {Wu}}, \bibinfo {author}
  {\bibfnamefont {X.}~\bibnamefont {Lu}}, \bibinfo {author} {\bibfnamefont
  {J.}~\bibnamefont {Choi}}, \bibinfo {author} {\bibfnamefont {K.}~\bibnamefont
  {Kim}}, \bibinfo {author} {\bibfnamefont {A.}~\bibnamefont {Rai}}, \bibinfo
  {author} {\bibfnamefont {D.~A.}\ \bibnamefont {Sanchez}}, \bibinfo {author}
  {\bibfnamefont {J.}~\bibnamefont {Quan}}, \bibinfo {author} {\bibfnamefont
  {A.}~\bibnamefont {Singh}}, \emph {et~al.},\ }\bibfield  {title} {\bibinfo
  {title} {Evidence for moir{\'e} excitons in van der waals heterostructures},\
  }\href@noop {} {\bibfield  {journal} {\bibinfo  {journal} {Nature}\ ,\
  \bibinfo {pages} {1}} (\bibinfo {year} {2019})}\BibitemShut {NoStop}%
\bibitem [{\citenamefont {Jones}\ \emph {et~al.}(2021)\citenamefont {Jones},
  \citenamefont {Muzzio}, \citenamefont {Pakdel}, \citenamefont {Biswas},
  \citenamefont {Curcio}, \citenamefont {Lanat{\`a}}, \citenamefont {Hofmann},
  \citenamefont {McCreary}, \citenamefont {Jonker}, \citenamefont {Watanabe}
  \emph {et~al.}}]{jones2021visualizing}%
  \BibitemOpen
  \bibfield  {author} {\bibinfo {author} {\bibfnamefont {A.~J.}\ \bibnamefont
  {Jones}}, \bibinfo {author} {\bibfnamefont {R.}~\bibnamefont {Muzzio}},
  \bibinfo {author} {\bibfnamefont {S.}~\bibnamefont {Pakdel}}, \bibinfo
  {author} {\bibfnamefont {D.}~\bibnamefont {Biswas}}, \bibinfo {author}
  {\bibfnamefont {D.}~\bibnamefont {Curcio}}, \bibinfo {author} {\bibfnamefont
  {N.}~\bibnamefont {Lanat{\`a}}}, \bibinfo {author} {\bibfnamefont
  {P.}~\bibnamefont {Hofmann}}, \bibinfo {author} {\bibfnamefont {K.~M.}\
  \bibnamefont {McCreary}}, \bibinfo {author} {\bibfnamefont {B.~T.}\
  \bibnamefont {Jonker}}, \bibinfo {author} {\bibfnamefont {K.}~\bibnamefont
  {Watanabe}}, \emph {et~al.},\ }\bibfield  {title} {\bibinfo {title}
  {Visualizing band structure hybridization and superlattice effects in twisted
  mos2/ws2 heterobilayers},\ }\href@noop {} {\bibfield  {journal} {\bibinfo
  {journal} {2D Materials}\ }\textbf {\bibinfo {volume} {9}},\ \bibinfo {pages}
  {015032} (\bibinfo {year} {2021})}\BibitemShut {NoStop}%
\bibitem [{\citenamefont {Wilson}\ and\ \citenamefont
  {et~al.}(2017)}]{Wilson2017}%
  \BibitemOpen
  \bibfield  {author} {\bibinfo {author} {\bibfnamefont {N.~R.}\ \bibnamefont
  {Wilson}}\ and\ \bibinfo {author} {\bibnamefont {et~al.}},\ }\bibfield
  {title} {\bibinfo {title} {{Determination of band offsets, hybridization, and
  exciton binding in 2D semiconductor heterostructures}},\ }\href@noop {}
  {\bibfield  {journal} {\bibinfo  {journal} {Science Advances}\ }\textbf
  {\bibinfo {volume} {3}},\ \bibinfo {pages} {1} (\bibinfo {year}
  {2017})}\BibitemShut {NoStop}%
\bibitem [{\citenamefont {Baek}\ \emph {et~al.}(2020)\citenamefont {Baek},
  \citenamefont {Brotons-Gisbert}, \citenamefont {Koong}, \citenamefont
  {Campbell}, \citenamefont {Rambach}, \citenamefont {Watanabe}, \citenamefont
  {Taniguchi},\ and\ \citenamefont {Gerardot}}]{baek2020highly}%
  \BibitemOpen
  \bibfield  {author} {\bibinfo {author} {\bibfnamefont {H.}~\bibnamefont
  {Baek}}, \bibinfo {author} {\bibfnamefont {M.}~\bibnamefont
  {Brotons-Gisbert}}, \bibinfo {author} {\bibfnamefont {Z.~X.}\ \bibnamefont
  {Koong}}, \bibinfo {author} {\bibfnamefont {A.}~\bibnamefont {Campbell}},
  \bibinfo {author} {\bibfnamefont {M.}~\bibnamefont {Rambach}}, \bibinfo
  {author} {\bibfnamefont {K.}~\bibnamefont {Watanabe}}, \bibinfo {author}
  {\bibfnamefont {T.}~\bibnamefont {Taniguchi}},\ and\ \bibinfo {author}
  {\bibfnamefont {B.~D.}\ \bibnamefont {Gerardot}},\ }\bibfield  {title}
  {\bibinfo {title} {Highly tunable quantum light from moir$\backslash$'e
  trapped excitons},\ }\href@noop {} {\bibfield  {journal} {\bibinfo  {journal}
  {arXiv preprint arXiv:2001.04305}\ } (\bibinfo {year} {2020})}\BibitemShut
  {NoStop}%
\bibitem [{\citenamefont {Tang}\ \emph {et~al.}(2020)\citenamefont {Tang},
  \citenamefont {Li}, \citenamefont {Li}, \citenamefont {Xu}, \citenamefont
  {Liu}, \citenamefont {Barmak}, \citenamefont {Watanabe}, \citenamefont
  {Taniguchi}, \citenamefont {MacDonald}, \citenamefont {Shan} \emph
  {et~al.}}]{tang2020simulation}%
  \BibitemOpen
  \bibfield  {author} {\bibinfo {author} {\bibfnamefont {Y.}~\bibnamefont
  {Tang}}, \bibinfo {author} {\bibfnamefont {L.}~\bibnamefont {Li}}, \bibinfo
  {author} {\bibfnamefont {T.}~\bibnamefont {Li}}, \bibinfo {author}
  {\bibfnamefont {Y.}~\bibnamefont {Xu}}, \bibinfo {author} {\bibfnamefont
  {S.}~\bibnamefont {Liu}}, \bibinfo {author} {\bibfnamefont {K.}~\bibnamefont
  {Barmak}}, \bibinfo {author} {\bibfnamefont {K.}~\bibnamefont {Watanabe}},
  \bibinfo {author} {\bibfnamefont {T.}~\bibnamefont {Taniguchi}}, \bibinfo
  {author} {\bibfnamefont {A.~H.}\ \bibnamefont {MacDonald}}, \bibinfo {author}
  {\bibfnamefont {J.}~\bibnamefont {Shan}}, \emph {et~al.},\ }\bibfield
  {title} {\bibinfo {title} {Simulation of hubbard model physics in wse2/ws2
  moir{\'e} superlattices},\ }\href@noop {} {\bibfield  {journal} {\bibinfo
  {journal} {Nature}\ }\textbf {\bibinfo {volume} {579}},\ \bibinfo {pages}
  {353} (\bibinfo {year} {2020})}\BibitemShut {NoStop}%
\bibitem [{\citenamefont {Xu}\ \emph {et~al.}(2022)\citenamefont {Xu},
  \citenamefont {Kang}, \citenamefont {Watanabe}, \citenamefont {Taniguchi},
  \citenamefont {Mak},\ and\ \citenamefont {Shan}}]{xu2022tunable}%
  \BibitemOpen
  \bibfield  {author} {\bibinfo {author} {\bibfnamefont {Y.}~\bibnamefont
  {Xu}}, \bibinfo {author} {\bibfnamefont {K.}~\bibnamefont {Kang}}, \bibinfo
  {author} {\bibfnamefont {K.}~\bibnamefont {Watanabe}}, \bibinfo {author}
  {\bibfnamefont {T.}~\bibnamefont {Taniguchi}}, \bibinfo {author}
  {\bibfnamefont {K.~F.}\ \bibnamefont {Mak}},\ and\ \bibinfo {author}
  {\bibfnamefont {J.}~\bibnamefont {Shan}},\ }\bibfield  {title} {\bibinfo
  {title} {A tunable bilayer hubbard model in twisted wse2},\ }\href@noop {}
  {\bibfield  {journal} {\bibinfo  {journal} {Nature Nanotechnology}\ }\textbf
  {\bibinfo {volume} {17}},\ \bibinfo {pages} {934} (\bibinfo {year}
  {2022})}\BibitemShut {NoStop}%
\bibitem [{\citenamefont {Shimazaki}\ \emph {et~al.}(2020)\citenamefont
  {Shimazaki}, \citenamefont {Schwartz}, \citenamefont {Watanabe},
  \citenamefont {Taniguchi}, \citenamefont {Kroner},\ and\ \citenamefont
  {Imamo{\u{g}}lu}}]{shimazaki2020strongly}%
  \BibitemOpen
  \bibfield  {author} {\bibinfo {author} {\bibfnamefont {Y.}~\bibnamefont
  {Shimazaki}}, \bibinfo {author} {\bibfnamefont {I.}~\bibnamefont {Schwartz}},
  \bibinfo {author} {\bibfnamefont {K.}~\bibnamefont {Watanabe}}, \bibinfo
  {author} {\bibfnamefont {T.}~\bibnamefont {Taniguchi}}, \bibinfo {author}
  {\bibfnamefont {M.}~\bibnamefont {Kroner}},\ and\ \bibinfo {author}
  {\bibfnamefont {A.}~\bibnamefont {Imamo{\u{g}}lu}},\ }\bibfield  {title}
  {\bibinfo {title} {Strongly correlated electrons and hybrid excitons in a
  moir{\'e} heterostructure},\ }\href@noop {} {\bibfield  {journal} {\bibinfo
  {journal} {Nature}\ }\textbf {\bibinfo {volume} {580}},\ \bibinfo {pages}
  {472} (\bibinfo {year} {2020})}\BibitemShut {NoStop}%
\bibitem [{\citenamefont {Rivera}\ \emph {et~al.}(2015)\citenamefont {Rivera},
  \citenamefont {Schaibley}, \citenamefont {Jones}, \citenamefont {Ross},
  \citenamefont {Wu}, \citenamefont {Aivazian}, \citenamefont {Klement},
  \citenamefont {Seyler}, \citenamefont {Clark}, \citenamefont {Ghimire} \emph
  {et~al.}}]{rivera2015observation}%
  \BibitemOpen
  \bibfield  {author} {\bibinfo {author} {\bibfnamefont {P.}~\bibnamefont
  {Rivera}}, \bibinfo {author} {\bibfnamefont {J.~R.}\ \bibnamefont
  {Schaibley}}, \bibinfo {author} {\bibfnamefont {A.~M.}\ \bibnamefont
  {Jones}}, \bibinfo {author} {\bibfnamefont {J.~S.}\ \bibnamefont {Ross}},
  \bibinfo {author} {\bibfnamefont {S.}~\bibnamefont {Wu}}, \bibinfo {author}
  {\bibfnamefont {G.}~\bibnamefont {Aivazian}}, \bibinfo {author}
  {\bibfnamefont {P.}~\bibnamefont {Klement}}, \bibinfo {author} {\bibfnamefont
  {K.}~\bibnamefont {Seyler}}, \bibinfo {author} {\bibfnamefont
  {G.}~\bibnamefont {Clark}}, \bibinfo {author} {\bibfnamefont {N.~J.}\
  \bibnamefont {Ghimire}}, \emph {et~al.},\ }\bibfield  {title} {\bibinfo
  {title} {Observation of long-lived interlayer excitons in monolayer mose
  2--wse 2 heterostructures},\ }\href@noop {} {\bibfield  {journal} {\bibinfo
  {journal} {Nature communications}\ }\textbf {\bibinfo {volume} {6}},\
  \bibinfo {pages} {1} (\bibinfo {year} {2015})}\BibitemShut {NoStop}%
\bibitem [{\citenamefont {Rivera}\ \emph {et~al.}(2018)\citenamefont {Rivera},
  \citenamefont {Yu}, \citenamefont {Seyler}, \citenamefont {Wilson},
  \citenamefont {Yao},\ and\ \citenamefont {Xu}}]{rivera2018interlayer}%
  \BibitemOpen
  \bibfield  {author} {\bibinfo {author} {\bibfnamefont {P.}~\bibnamefont
  {Rivera}}, \bibinfo {author} {\bibfnamefont {H.}~\bibnamefont {Yu}}, \bibinfo
  {author} {\bibfnamefont {K.~L.}\ \bibnamefont {Seyler}}, \bibinfo {author}
  {\bibfnamefont {N.~P.}\ \bibnamefont {Wilson}}, \bibinfo {author}
  {\bibfnamefont {W.}~\bibnamefont {Yao}},\ and\ \bibinfo {author}
  {\bibfnamefont {X.}~\bibnamefont {Xu}},\ }\bibfield  {title} {\bibinfo
  {title} {Interlayer valley excitons in heterobilayers of transition metal
  dichalcogenides},\ }\href@noop {} {\bibfield  {journal} {\bibinfo  {journal}
  {Nature nanotechnology}\ ,\ \bibinfo {pages} {1}} (\bibinfo {year}
  {2018})}\BibitemShut {NoStop}%
\bibitem [{\citenamefont {Jiang}\ \emph {et~al.}(2021)\citenamefont {Jiang},
  \citenamefont {Chen}, \citenamefont {Zheng}, \citenamefont {Zheng},\ and\
  \citenamefont {Pan}}]{jiang2021interlayer}%
  \BibitemOpen
  \bibfield  {author} {\bibinfo {author} {\bibfnamefont {Y.}~\bibnamefont
  {Jiang}}, \bibinfo {author} {\bibfnamefont {S.}~\bibnamefont {Chen}},
  \bibinfo {author} {\bibfnamefont {W.}~\bibnamefont {Zheng}}, \bibinfo
  {author} {\bibfnamefont {B.}~\bibnamefont {Zheng}},\ and\ \bibinfo {author}
  {\bibfnamefont {A.}~\bibnamefont {Pan}},\ }\bibfield  {title} {\bibinfo
  {title} {Interlayer exciton formation, relaxation, and transport in tmd van
  der waals heterostructures},\ }\href@noop {} {\bibfield  {journal} {\bibinfo
  {journal} {Light: Science \& Applications}\ }\textbf {\bibinfo {volume}
  {10}},\ \bibinfo {pages} {72} (\bibinfo {year} {2021})}\BibitemShut {NoStop}%
\bibitem [{\citenamefont {Adachi}(2009)}]{adachi2009properties}%
  \BibitemOpen
  \bibfield  {author} {\bibinfo {author} {\bibfnamefont {S.}~\bibnamefont
  {Adachi}},\ }\href@noop {} {\emph {\bibinfo {title} {Properties of
  semiconductor alloys: group-IV, III-V and II-VI semiconductors}}},\
  Vol.~\bibinfo {volume} {28}\ (\bibinfo  {publisher} {John Wiley \& Sons},\
  \bibinfo {year} {2009})\BibitemShut {NoStop}%
\bibitem [{\citenamefont {Xie}(2015)}]{xie2015two}%
  \BibitemOpen
  \bibfield  {author} {\bibinfo {author} {\bibfnamefont {L.}~\bibnamefont
  {Xie}},\ }\bibfield  {title} {\bibinfo {title} {Two-dimensional transition
  metal dichalcogenide alloys: preparation, characterization and
  applications},\ }\href@noop {} {\bibfield  {journal} {\bibinfo  {journal}
  {Nanoscale}\ }\textbf {\bibinfo {volume} {7}},\ \bibinfo {pages} {18392}
  (\bibinfo {year} {2015})}\BibitemShut {NoStop}%
\bibitem [{\citenamefont {Wang}\ and\ \citenamefont
  {et~al.}(2015)}]{Wang2015a}%
  \BibitemOpen
  \bibfield  {author} {\bibinfo {author} {\bibfnamefont {G.}~\bibnamefont
  {Wang}}\ and\ \bibinfo {author} {\bibnamefont {et~al.}},\ }\bibfield  {title}
  {\bibinfo {title} {{Spin-orbit engineering in transition metal dichalcogenide
  alloy monolayers}},\ }\href@noop {} {\bibfield  {journal} {\bibinfo
  {journal} {Nature Communications}\ }\textbf {\bibinfo {volume} {6}},\
  \bibinfo {pages} {1} (\bibinfo {year} {2015})}\BibitemShut {NoStop}%
\bibitem [{\citenamefont {Ye}\ \emph {et~al.}(2017)\citenamefont {Ye},
  \citenamefont {Niu}, \citenamefont {Li}, \citenamefont {Li},\ and\
  \citenamefont {Zhang}}]{ye2017exciton}%
  \BibitemOpen
  \bibfield  {author} {\bibinfo {author} {\bibfnamefont {J.}~\bibnamefont
  {Ye}}, \bibinfo {author} {\bibfnamefont {B.}~\bibnamefont {Niu}}, \bibinfo
  {author} {\bibfnamefont {Y.}~\bibnamefont {Li}}, \bibinfo {author}
  {\bibfnamefont {T.}~\bibnamefont {Li}},\ and\ \bibinfo {author}
  {\bibfnamefont {X.}~\bibnamefont {Zhang}},\ }\bibfield  {title} {\bibinfo
  {title} {Exciton valley dynamics in monolayer mo1-xwxse2 (x= 0, 0.5, 1)},\
  }\href@noop {} {\bibfield  {journal} {\bibinfo  {journal} {Applied Physics
  Letters}\ }\textbf {\bibinfo {volume} {111}},\ \bibinfo {pages} {152106}
  (\bibinfo {year} {2017})}\BibitemShut {NoStop}%
\bibitem [{\citenamefont {Li}\ \emph {et~al.}(2020)\citenamefont {Li},
  \citenamefont {Zheng}, \citenamefont {Ma}, \citenamefont {Zhao},
  \citenamefont {Jiang}, \citenamefont {Ouyang}, \citenamefont {Zheng},
  \citenamefont {Fu}, \citenamefont {Fan}, \citenamefont {Zheng} \emph
  {et~al.}}]{li2020wavelength}%
  \BibitemOpen
  \bibfield  {author} {\bibinfo {author} {\bibfnamefont {L.}~\bibnamefont
  {Li}}, \bibinfo {author} {\bibfnamefont {W.}~\bibnamefont {Zheng}}, \bibinfo
  {author} {\bibfnamefont {C.}~\bibnamefont {Ma}}, \bibinfo {author}
  {\bibfnamefont {H.}~\bibnamefont {Zhao}}, \bibinfo {author} {\bibfnamefont
  {F.}~\bibnamefont {Jiang}}, \bibinfo {author} {\bibfnamefont
  {Y.}~\bibnamefont {Ouyang}}, \bibinfo {author} {\bibfnamefont
  {B.}~\bibnamefont {Zheng}}, \bibinfo {author} {\bibfnamefont
  {X.}~\bibnamefont {Fu}}, \bibinfo {author} {\bibfnamefont {P.}~\bibnamefont
  {Fan}}, \bibinfo {author} {\bibfnamefont {M.}~\bibnamefont {Zheng}}, \emph
  {et~al.},\ }\bibfield  {title} {\bibinfo {title} {Wavelength tunable
  interlayer exciton emission at near-infrared region in van der waals
  semiconductor heterostructures},\ }\href@noop {} {\bibfield  {journal}
  {\bibinfo  {journal} {Nano Letters}\ } (\bibinfo {year} {2020})}\BibitemShut
  {NoStop}%
\bibitem [{\citenamefont {Zi}\ \emph {et~al.}(2019)\citenamefont {Zi},
  \citenamefont {Li}, \citenamefont {Niu}, \citenamefont {Wang}, \citenamefont
  {Cho},\ and\ \citenamefont {Jia}}]{zi2019reversible}%
  \BibitemOpen
  \bibfield  {author} {\bibinfo {author} {\bibfnamefont {Y.}~\bibnamefont
  {Zi}}, \bibinfo {author} {\bibfnamefont {C.}~\bibnamefont {Li}}, \bibinfo
  {author} {\bibfnamefont {C.}~\bibnamefont {Niu}}, \bibinfo {author}
  {\bibfnamefont {F.}~\bibnamefont {Wang}}, \bibinfo {author} {\bibfnamefont
  {J.-H.}\ \bibnamefont {Cho}},\ and\ \bibinfo {author} {\bibfnamefont
  {Y.}~\bibnamefont {Jia}},\ }\bibfield  {title} {\bibinfo {title} {Reversible
  direct-indirect band transition in alloying tmds heterostructures via band
  engineering},\ }\href@noop {} {\bibfield  {journal} {\bibinfo  {journal}
  {Journal of Physics: Condensed Matter}\ }\textbf {\bibinfo {volume} {31}},\
  \bibinfo {pages} {435503} (\bibinfo {year} {2019})}\BibitemShut {NoStop}%
\bibitem [{\citenamefont {Alexeev}\ and\ \citenamefont
  {et~al.}(2017)}]{Alexeev2017}%
  \BibitemOpen
  \bibfield  {author} {\bibinfo {author} {\bibfnamefont {E.~M.}\ \bibnamefont
  {Alexeev}}\ and\ \bibinfo {author} {\bibnamefont {et~al.}},\ }\bibfield
  {title} {\bibinfo {title} {{Imaging of interlayer coupling in van der Waals
  heterostructures using a bright-field optical microscope}},\ }\href@noop {}
  {\bibfield  {journal} {\bibinfo  {journal} {Nano Letters}\ }\textbf {\bibinfo
  {volume} {17}},\ \bibinfo {pages} {5342} (\bibinfo {year}
  {2017})}\BibitemShut {NoStop}%
\bibitem [{\citenamefont {Hong}\ and\ \citenamefont {et~al.}(2014)}]{Hong2014}%
  \BibitemOpen
  \bibfield  {author} {\bibinfo {author} {\bibfnamefont {X.}~\bibnamefont
  {Hong}}\ and\ \bibinfo {author} {\bibnamefont {et~al.}},\ }\bibfield  {title}
  {\bibinfo {title} {{Ultrafast charge transfer in atomically thin
  MoS\textsubscript{2}/WS\textsubscript{2} heterostructures}},\ }\href@noop {}
  {\bibfield  {journal} {\bibinfo  {journal} {Nature Nanotechnology}\ }\textbf
  {\bibinfo {volume} {9}},\ \bibinfo {pages} {1} (\bibinfo {year}
  {2014})}\BibitemShut {NoStop}%
\bibitem [{\citenamefont {Rivera}\ and\ \citenamefont
  {et~al.}(2015)}]{Rivera2015}%
  \BibitemOpen
  \bibfield  {author} {\bibinfo {author} {\bibfnamefont {P.}~\bibnamefont
  {Rivera}}\ and\ \bibinfo {author} {\bibnamefont {et~al.}},\ }\bibfield
  {title} {\bibinfo {title} {{Observation of long-lived interlayer excitons in
  monolayer MoSe\textsubscript{2}-WSe\textsubscript{2} heterostructures}},\
  }\href@noop {} {\bibfield  {journal} {\bibinfo  {journal} {Nature
  communications}\ }\textbf {\bibinfo {volume} {6}},\ \bibinfo {pages} {6242}
  (\bibinfo {year} {2015})}\BibitemShut {NoStop}%
\bibitem [{\citenamefont {Kopaczek}\ \emph {et~al.}(2021)\citenamefont
  {Kopaczek}, \citenamefont {Wozniak}, \citenamefont {Tamulewicz-Szwajkowska},
  \citenamefont {Zelewski}, \citenamefont {Serafinczuk}, \citenamefont
  {Scharoch},\ and\ \citenamefont {Kudrawiec}}]{kopaczek2021experimental}%
  \BibitemOpen
  \bibfield  {author} {\bibinfo {author} {\bibfnamefont {J.}~\bibnamefont
  {Kopaczek}}, \bibinfo {author} {\bibfnamefont {T.}~\bibnamefont {Wozniak}},
  \bibinfo {author} {\bibfnamefont {M.}~\bibnamefont {Tamulewicz-Szwajkowska}},
  \bibinfo {author} {\bibfnamefont {S.~J.}\ \bibnamefont {Zelewski}}, \bibinfo
  {author} {\bibfnamefont {J.}~\bibnamefont {Serafinczuk}}, \bibinfo {author}
  {\bibfnamefont {P.}~\bibnamefont {Scharoch}},\ and\ \bibinfo {author}
  {\bibfnamefont {R.}~\bibnamefont {Kudrawiec}},\ }\bibfield  {title} {\bibinfo
  {title} {Experimental and theoretical studies of the electronic band
  structure of bulk and atomically thin mo1--x w x se2 alloys},\ }\href@noop {}
  {\bibfield  {journal} {\bibinfo  {journal} {ACS omega}\ }\textbf {\bibinfo
  {volume} {6}},\ \bibinfo {pages} {19893} (\bibinfo {year}
  {2021})}\BibitemShut {NoStop}%
\bibitem [{\citenamefont {Denton}\ and\ \citenamefont
  {Ashcroft}(1991)}]{denton1991vegard}%
  \BibitemOpen
  \bibfield  {author} {\bibinfo {author} {\bibfnamefont {A.~R.}\ \bibnamefont
  {Denton}}\ and\ \bibinfo {author} {\bibfnamefont {N.~W.}\ \bibnamefont
  {Ashcroft}},\ }\bibfield  {title} {\bibinfo {title} {Vegard?s law},\
  }\href@noop {} {\bibfield  {journal} {\bibinfo  {journal} {Physical review
  A}\ }\textbf {\bibinfo {volume} {43}},\ \bibinfo {pages} {3161} (\bibinfo
  {year} {1991})}\BibitemShut {NoStop}%
\bibitem [{\citenamefont {Tongay}\ and\ \citenamefont
  {et~al.}(2014)}]{Tongay2014}%
  \BibitemOpen
  \bibfield  {author} {\bibinfo {author} {\bibfnamefont {S.}~\bibnamefont
  {Tongay}}\ and\ \bibinfo {author} {\bibnamefont {et~al.}},\ }\bibfield
  {title} {\bibinfo {title} {{Two-dimensional semiconductor alloys: Monolayer
  Mo\textsubscript{1-x}W\textsubscript{x}Se\textsubscript{2}}},\ }\href@noop {}
  {\bibfield  {journal} {\bibinfo  {journal} {Applied Physics Letters}\
  }\textbf {\bibinfo {volume} {104}},\ \bibinfo {pages} {1} (\bibinfo {year}
  {2014})}\BibitemShut {NoStop}%
\bibitem [{\citenamefont {Zhang}\ and\ \citenamefont
  {et~al.}(2014)}]{Zhang2014}%
  \BibitemOpen
  \bibfield  {author} {\bibinfo {author} {\bibfnamefont {M.}~\bibnamefont
  {Zhang}}\ and\ \bibinfo {author} {\bibnamefont {et~al.}},\ }\bibfield
  {title} {\bibinfo {title} {{Two-Dimensional Molybdenum Tungsten Diselenide
  Alloys: Photoluminescence, Raman Scattering, and Electrical Transport}},\
  }\href@noop {} {\bibfield  {journal} {\bibinfo  {journal} {ACS Nano}\
  }\textbf {\bibinfo {volume} {8}},\ \bibinfo {pages} {7130} (\bibinfo {year}
  {2014})}\BibitemShut {NoStop}%
\bibitem [{\citenamefont {Molas}\ \emph {et~al.}(2017)\citenamefont {Molas},
  \citenamefont {Nogajewski}, \citenamefont {Slobodeniuk}, \citenamefont
  {Binder}, \citenamefont {Bartos},\ and\ \citenamefont
  {Potemski}}]{molas2017optical}%
  \BibitemOpen
  \bibfield  {author} {\bibinfo {author} {\bibfnamefont {M.~R.}\ \bibnamefont
  {Molas}}, \bibinfo {author} {\bibfnamefont {K.}~\bibnamefont {Nogajewski}},
  \bibinfo {author} {\bibfnamefont {A.~O.}\ \bibnamefont {Slobodeniuk}},
  \bibinfo {author} {\bibfnamefont {J.}~\bibnamefont {Binder}}, \bibinfo
  {author} {\bibfnamefont {M.}~\bibnamefont {Bartos}},\ and\ \bibinfo {author}
  {\bibfnamefont {M.}~\bibnamefont {Potemski}},\ }\bibfield  {title} {\bibinfo
  {title} {The optical response of monolayer, few-layer and bulk tungsten
  disulfide},\ }\href@noop {} {\bibfield  {journal} {\bibinfo  {journal}
  {Nanoscale}\ }\textbf {\bibinfo {volume} {9}},\ \bibinfo {pages} {13128}
  (\bibinfo {year} {2017})}\BibitemShut {NoStop}%
\bibitem [{\citenamefont {Miller}\ \emph {et~al.}(2017)\citenamefont {Miller},
  \citenamefont {Steinhoff}, \citenamefont {Pano}, \citenamefont {Klein},
  \citenamefont {Jahnke}, \citenamefont {Holleitner},\ and\ \citenamefont
  {Wurstbauer}}]{miller2017long}%
  \BibitemOpen
  \bibfield  {author} {\bibinfo {author} {\bibfnamefont {B.}~\bibnamefont
  {Miller}}, \bibinfo {author} {\bibfnamefont {A.}~\bibnamefont {Steinhoff}},
  \bibinfo {author} {\bibfnamefont {B.}~\bibnamefont {Pano}}, \bibinfo {author}
  {\bibfnamefont {J.}~\bibnamefont {Klein}}, \bibinfo {author} {\bibfnamefont
  {F.}~\bibnamefont {Jahnke}}, \bibinfo {author} {\bibfnamefont
  {A.}~\bibnamefont {Holleitner}},\ and\ \bibinfo {author} {\bibfnamefont
  {U.}~\bibnamefont {Wurstbauer}},\ }\bibfield  {title} {\bibinfo {title}
  {Long-lived direct and indirect interlayer excitons in van der waals
  heterostructures},\ }\href@noop {} {\bibfield  {journal} {\bibinfo  {journal}
  {Nano letters}\ }\textbf {\bibinfo {volume} {17}},\ \bibinfo {pages} {5229}
  (\bibinfo {year} {2017})}\BibitemShut {NoStop}%
\bibitem [{\citenamefont {Barr{\'e}}\ \emph {et~al.}(2022)\citenamefont
  {Barr{\'e}}, \citenamefont {Karni}, \citenamefont {Liu}, \citenamefont
  {O’Beirne}, \citenamefont {Chen}, \citenamefont {Ribeiro}, \citenamefont
  {Yu}, \citenamefont {Kim}, \citenamefont {Watanabe}, \citenamefont
  {Taniguchi} \emph {et~al.}}]{barre2022optical}%
  \BibitemOpen
  \bibfield  {author} {\bibinfo {author} {\bibfnamefont {E.}~\bibnamefont
  {Barr{\'e}}}, \bibinfo {author} {\bibfnamefont {O.}~\bibnamefont {Karni}},
  \bibinfo {author} {\bibfnamefont {E.}~\bibnamefont {Liu}}, \bibinfo {author}
  {\bibfnamefont {A.~L.}\ \bibnamefont {O’Beirne}}, \bibinfo {author}
  {\bibfnamefont {X.}~\bibnamefont {Chen}}, \bibinfo {author} {\bibfnamefont
  {H.~B.}\ \bibnamefont {Ribeiro}}, \bibinfo {author} {\bibfnamefont
  {L.}~\bibnamefont {Yu}}, \bibinfo {author} {\bibfnamefont {B.}~\bibnamefont
  {Kim}}, \bibinfo {author} {\bibfnamefont {K.}~\bibnamefont {Watanabe}},
  \bibinfo {author} {\bibfnamefont {T.}~\bibnamefont {Taniguchi}}, \emph
  {et~al.},\ }\bibfield  {title} {\bibinfo {title} {Optical absorption of
  interlayer excitons in transition-metal dichalcogenide heterostructures},\
  }\href@noop {} {\bibfield  {journal} {\bibinfo  {journal} {Science}\ }\textbf
  {\bibinfo {volume} {376}},\ \bibinfo {pages} {406} (\bibinfo {year}
  {2022})}\BibitemShut {NoStop}%
\bibitem [{\citenamefont {Lindlau}\ and\ \citenamefont
  {et~al.}(2018)}]{Lindlau2018}%
  \BibitemOpen
  \bibfield  {author} {\bibinfo {author} {\bibfnamefont {J.}~\bibnamefont
  {Lindlau}}\ and\ \bibinfo {author} {\bibnamefont {et~al.}},\ }\bibfield
  {title} {\bibinfo {title} {{The role of momentum-dark excitons in the
  elementary optical response of bilayer WSe\textsubscript{2}}},\ }\href@noop
  {} {\bibfield  {journal} {\bibinfo  {journal} {Nature Communications}\
  }\textbf {\bibinfo {volume} {9}},\ \bibinfo {pages} {1} (\bibinfo {year}
  {2018})}\BibitemShut {NoStop}%
\bibitem [{\citenamefont {Ovesen}\ \emph {et~al.}(2019)\citenamefont {Ovesen},
  \citenamefont {Brem}, \citenamefont {Linder{\"a}lv}, \citenamefont {Kuisma},
  \citenamefont {Korn}, \citenamefont {Erhart}, \citenamefont {Selig},\ and\
  \citenamefont {Malic}}]{ovesen2019interlayer}%
  \BibitemOpen
  \bibfield  {author} {\bibinfo {author} {\bibfnamefont {S.}~\bibnamefont
  {Ovesen}}, \bibinfo {author} {\bibfnamefont {S.}~\bibnamefont {Brem}},
  \bibinfo {author} {\bibfnamefont {C.}~\bibnamefont {Linder{\"a}lv}}, \bibinfo
  {author} {\bibfnamefont {M.}~\bibnamefont {Kuisma}}, \bibinfo {author}
  {\bibfnamefont {T.}~\bibnamefont {Korn}}, \bibinfo {author} {\bibfnamefont
  {P.}~\bibnamefont {Erhart}}, \bibinfo {author} {\bibfnamefont
  {M.}~\bibnamefont {Selig}},\ and\ \bibinfo {author} {\bibfnamefont
  {E.}~\bibnamefont {Malic}},\ }\bibfield  {title} {\bibinfo {title}
  {Interlayer exciton dynamics in van der waals heterostructures},\ }\href@noop
  {} {\bibfield  {journal} {\bibinfo  {journal} {Communications Physics}\
  }\textbf {\bibinfo {volume} {2}},\ \bibinfo {pages} {23} (\bibinfo {year}
  {2019})}\BibitemShut {NoStop}%
\bibitem [{\citenamefont {Wang}\ and\ \citenamefont
  {et~al.}(2014)}]{Wang2014a}%
  \BibitemOpen
  \bibfield  {author} {\bibinfo {author} {\bibfnamefont {G.}~\bibnamefont
  {Wang}}\ and\ \bibinfo {author} {\bibnamefont {et~al.}},\ }\bibfield  {title}
  {\bibinfo {title} {{Exciton dynamics in WSe\textsubscript{2} bilayers}},\
  }\href@noop {} {\bibfield  {journal} {\bibinfo  {journal} {Applied Physics
  Letters}\ }\textbf {\bibinfo {volume} {105}},\ \bibinfo {pages} {1} (\bibinfo
  {year} {2014})}\BibitemShut {NoStop}%
\bibitem [{\citenamefont {Wang}\ and\ \citenamefont {et~al.}(2018)}]{Wang2018}%
  \BibitemOpen
  \bibfield  {author} {\bibinfo {author} {\bibfnamefont {Z.}~\bibnamefont
  {Wang}}\ and\ \bibinfo {author} {\bibnamefont {et~al.}},\ }\bibfield  {title}
  {\bibinfo {title} {{Electrical Tuning of Interlayer Exciton Gases in
  WSe\textsubscript{2} Bilayers}},\ }\href@noop {} {\bibfield  {journal}
  {\bibinfo  {journal} {Nano Letters}\ }\textbf {\bibinfo {volume} {18}},\
  \bibinfo {pages} {137} (\bibinfo {year} {2018})}\BibitemShut {NoStop}%
\bibitem [{\citenamefont {Wang}\ \emph {et~al.}(2014)\citenamefont {Wang},
  \citenamefont {Marie}, \citenamefont {Bouet}, \citenamefont {Vidal},
  \citenamefont {Balocchi}, \citenamefont {Amand}, \citenamefont {Lagarde},\
  and\ \citenamefont {Urbaszek}}]{wang2014exciton}%
  \BibitemOpen
  \bibfield  {author} {\bibinfo {author} {\bibfnamefont {G.}~\bibnamefont
  {Wang}}, \bibinfo {author} {\bibfnamefont {X.}~\bibnamefont {Marie}},
  \bibinfo {author} {\bibfnamefont {L.}~\bibnamefont {Bouet}}, \bibinfo
  {author} {\bibfnamefont {M.}~\bibnamefont {Vidal}}, \bibinfo {author}
  {\bibfnamefont {A.}~\bibnamefont {Balocchi}}, \bibinfo {author}
  {\bibfnamefont {T.}~\bibnamefont {Amand}}, \bibinfo {author} {\bibfnamefont
  {D.}~\bibnamefont {Lagarde}},\ and\ \bibinfo {author} {\bibfnamefont
  {B.}~\bibnamefont {Urbaszek}},\ }\bibfield  {title} {\bibinfo {title}
  {Exciton dynamics in wse2 bilayers},\ }\href@noop {} {\bibfield  {journal}
  {\bibinfo  {journal} {Applied Physics Letters}\ }\textbf {\bibinfo {volume}
  {105}},\ \bibinfo {pages} {182105} (\bibinfo {year} {2014})}\BibitemShut
  {NoStop}%
\bibitem [{\citenamefont {Sun}\ \emph {et~al.}(2014)\citenamefont {Sun},
  \citenamefont {Rao}, \citenamefont {Reider}, \citenamefont {Chen},
  \citenamefont {You}, \citenamefont {Bre?zin}, \citenamefont {Harutyunyan},\
  and\ \citenamefont {Heinz}}]{sun2014observation}%
  \BibitemOpen
  \bibfield  {author} {\bibinfo {author} {\bibfnamefont {D.}~\bibnamefont
  {Sun}}, \bibinfo {author} {\bibfnamefont {Y.}~\bibnamefont {Rao}}, \bibinfo
  {author} {\bibfnamefont {G.~A.}\ \bibnamefont {Reider}}, \bibinfo {author}
  {\bibfnamefont {G.}~\bibnamefont {Chen}}, \bibinfo {author} {\bibfnamefont
  {Y.}~\bibnamefont {You}}, \bibinfo {author} {\bibfnamefont {L.}~\bibnamefont
  {Bre?zin}}, \bibinfo {author} {\bibfnamefont {A.~R.}\ \bibnamefont
  {Harutyunyan}},\ and\ \bibinfo {author} {\bibfnamefont {T.~F.}\ \bibnamefont
  {Heinz}},\ }\bibfield  {title} {\bibinfo {title} {Observation of rapid
  exciton--exciton annihilation in monolayer molybdenum disulfide},\
  }\href@noop {} {\bibfield  {journal} {\bibinfo  {journal} {Nano letters}\
  }\textbf {\bibinfo {volume} {14}},\ \bibinfo {pages} {5625} (\bibinfo {year}
  {2014})}\BibitemShut {NoStop}%
\bibitem [{\citenamefont {Kravtsov}\ \emph {et~al.}(2021)\citenamefont
  {Kravtsov}, \citenamefont {Liubomirov}, \citenamefont {Cherbunin},
  \citenamefont {Catanzaro}, \citenamefont {Genco}, \citenamefont {Gillard},
  \citenamefont {Alexeev}, \citenamefont {Ivanova}, \citenamefont {Khestanova},
  \citenamefont {Shelykh} \emph {et~al.}}]{kravtsov2021spin}%
  \BibitemOpen
  \bibfield  {author} {\bibinfo {author} {\bibfnamefont {V.}~\bibnamefont
  {Kravtsov}}, \bibinfo {author} {\bibfnamefont {A.~D.}\ \bibnamefont
  {Liubomirov}}, \bibinfo {author} {\bibfnamefont {R.~V.}\ \bibnamefont
  {Cherbunin}}, \bibinfo {author} {\bibfnamefont {A.}~\bibnamefont
  {Catanzaro}}, \bibinfo {author} {\bibfnamefont {A.}~\bibnamefont {Genco}},
  \bibinfo {author} {\bibfnamefont {D.}~\bibnamefont {Gillard}}, \bibinfo
  {author} {\bibfnamefont {E.~M.}\ \bibnamefont {Alexeev}}, \bibinfo {author}
  {\bibfnamefont {T.}~\bibnamefont {Ivanova}}, \bibinfo {author} {\bibfnamefont
  {E.}~\bibnamefont {Khestanova}}, \bibinfo {author} {\bibfnamefont {I.~A.}\
  \bibnamefont {Shelykh}}, \emph {et~al.},\ }\bibfield  {title} {\bibinfo
  {title} {Spin--valley dynamics in alloy-based transition metal dichalcogenide
  heterobilayers},\ }\href@noop {} {\bibfield  {journal} {\bibinfo  {journal}
  {2D Materials}\ }\textbf {\bibinfo {volume} {8}},\ \bibinfo {pages} {025011}
  (\bibinfo {year} {2021})}\BibitemShut {NoStop}%
\bibitem [{\citenamefont {Magorrian}\ \emph {et~al.}(2021)\citenamefont
  {Magorrian}, \citenamefont {Enaldiev}, \citenamefont {Z{\'o}lyomi},
  \citenamefont {Ferreira}, \citenamefont {Fal'ko},\ and\ \citenamefont
  {Ruiz-Tijerina}}]{magorrian2021multifaceted}%
  \BibitemOpen
  \bibfield  {author} {\bibinfo {author} {\bibfnamefont {S.}~\bibnamefont
  {Magorrian}}, \bibinfo {author} {\bibfnamefont {V.}~\bibnamefont {Enaldiev}},
  \bibinfo {author} {\bibfnamefont {V.}~\bibnamefont {Z{\'o}lyomi}}, \bibinfo
  {author} {\bibfnamefont {F.}~\bibnamefont {Ferreira}}, \bibinfo {author}
  {\bibfnamefont {V.~I.}\ \bibnamefont {Fal'ko}},\ and\ \bibinfo {author}
  {\bibfnamefont {D.~A.}\ \bibnamefont {Ruiz-Tijerina}},\ }\bibfield  {title}
  {\bibinfo {title} {Multifaceted moir{\'e} superlattice physics in twisted wse
  2 bilayers},\ }\href@noop {} {\bibfield  {journal} {\bibinfo  {journal}
  {Physical Review B}\ }\textbf {\bibinfo {volume} {104}},\ \bibinfo {pages}
  {125440} (\bibinfo {year} {2021})}\BibitemShut {NoStop}%
\end{thebibliography}%

\end{document}